\documentclass[twocolumn]{aastex63}
\usepackage{amsmath}
\usepackage{graphicx}
\usepackage{subfigure}
\usepackage{multirow,array}
\usepackage{lineno}

\shorttitle{Polarization in X-ray Bursts}
\shortauthors{Zhong et al.}

\begin{document}
\title{Identifying the Origin of FRB-associated X-ray Bursts with X-ray Polarization}
\author[0000-0002-1766-6947]{Shu-Qing Zhong}
\affil{School of Science, Guangxi University of Science and Technology, Liuzhou 545006, People’s Republic of China}
\affil{Department of Astronomy, University of Science and Technology of China, Hefei 230026, People’s Republic of China; daizg@ustc.edu.cn}
\affil{School of Astronomy and Space Science, University of Science and Technology of China, Hefei 230026, People’s Republic of China}
\author[0000-0002-8391-5980]{Long Li}
\affil{Department of Physics, School of Physics and Materials Science, Nanchang University, Nanchang 330031, People’s Republic of China}
\affil{Department of Astronomy, University of Science and Technology of China, Hefei 230026, People’s Republic of China; daizg@ustc.edu.cn}
\affil{School of Astronomy and Space Science, University of Science and Technology of China, Hefei 230026, People’s Republic of China}
\author{Biao Zhang}
\affil{Department of Astronomy, University of Science and Technology of China, Hefei 230026, People’s Republic of China; daizg@ustc.edu.cn}
\affil{School of Astronomy and Space Science, University of Science and Technology of China, Hefei 230026, People’s Republic of China}
\author[0000-0002-7835-8585]{Zi-Gao Dai}
\affil{Department of Astronomy, University of Science and Technology of China, Hefei 230026, People’s Republic of China; daizg@ustc.edu.cn}
\affil{School of Astronomy and Space Science, University of Science and Technology of China, Hefei 230026, People’s Republic of China}

\begin{abstract}
The origin of extraordinary X-ray burst (XRB) associated with a fast radio burst (FRB) like FRB 20200428D is still unclear, 
though several models such as the emission of a trapped fireball modified by resonant cyclotron scattering, 
the outflow from a polar trapped-expanding fireball, and the synchrotron radiation of a far-away relativistic shock, 
have been proposed.
To determine which model is true, we study possible X-ray polarization signature for each model, inspired by the importance of radio polarization in identifying FRB origin.
We first numerically simulate or calculate the XRB spectrum for each model and fit it to the observed data, then compute the corresponding polarization signal based on the fit.
We find that these three models predict different polarization patterns in terms of phase/time and energy variations. 
The differences can be used to test the models with future X-ray polarization observations.
\end{abstract}

\keywords{Magnetars (992); Polarimetry(1278); Radio bursts (1339); X-ray bursts (1814); Radio transient sources(2008)}

\section{Introduction}
\label{sec:introduction}
Fast radio bursts (FRBs)
are millisecond cosmological radio flashes with particular characteristics \citep{lori07}, e.g., much higher luminosity and more extreme
brightness temperature compared with pulsar radio emission and radio transients from Galactic magnetars \citep[see reviews][]{cor19,pet19,pet22,zhang20,zhang23,xiao21}.
Although the physical origin of FRBs remains an open question, it is widely accepted that at least some of them are from magnetars, 
thanks to the discovery of FRB 20200428D \citep{chime20b,boch20}
and its associated X-ray burst \citep[XRB;][]{mere20,lick21,rid21,tava21} both from SGR 1935+2154 (abbr., SGR 1935).

In regard to the radiation mechanism of FRBs, two classes of models are commonly discussed.
One class are relative to coherent radiations invoking a magnetar magnetosphere, such as
coherent curvature radiation \citep{kumar17,yang18,kumar20,dai20},
coherent inverse Compton scattering \citep{zhang22},
collective plasma radiation due to nonstationary pair plasma discharges \citep{wad19,phi20,yang21},
and magnetic reconnection in a current sheet of magnetar wind beyond the light cylinder \citep{lyu20,mah22}.
The other class involve a synchrotron maser radiation from decelerating
relativistic blast waves far outside magnetar magnetosphere \citep[][for different upstream media]{lyu14,bel17,met19}.

When taking into account a possible XRB association for FRBs like the
FRB-associated XRB (FXRB) accompanying FRB 20200428D \citep{mere20,lick21,rid21,tava21},
one should propose mechanism models, either within or far outside magnetosphere, those can interpret the simultaneous generation
of an FRB and its FXRB, as already done by many authors \citep[e.g.,][]{lu20,kat20,ioka20,yuan20,yang21,met19,mar20}.
In those models within magnetosphere, 
the FXRB is thought to be initially involved with a trapped fireball produced by an abrupt magnetic energy dissipation due to a crustal deformation or fracture
\citep[e.g.,][]{lu20,ioka20,yang21} or with a magnetic reconnection event \citep{yuan20,xie23}. 
While in the models far outside magnetosphere, the FXRB is usually believed to be the incoherent synchrotron radiation from hot electrons heated by the shock 
due to the collision between relativistic ejecta and upstream medium \citep{lyu14,met19}.

Even though the FRB radiation mechanism is still being debated, more and more observations, especially in radio polarization such as the rapid and diverse
polarization angle swings in FRB 20180301A \citep{luo20} and circular polarization in FRB 20201124A \citep{xu22}, 
favor the models in magnetosphere. Similarly, the origin of an FXRB is a debate as well. 
To identify the origin of an FXRB, we intend to study its possible X-ray polarization properties and expect future X-ray polarization signal observations in this work.

The structure of the paper is organized as follows. 
In Section \ref{sec:extraordinary}, we illuminate the extraordinary features of an FXRB like the one accompanying FRB 20200428D 
and its plausible theoretical explanations. The X-ray polarization properties in different theoretical explanations are explored in Sections \ref{sec:rcs}-\ref{sec:synchrotron}.
The summary and discussion are finally presented in Section \ref{sec:summary}.

\begin{figure*}
	\centering
	\includegraphics[width=1.0\textwidth]{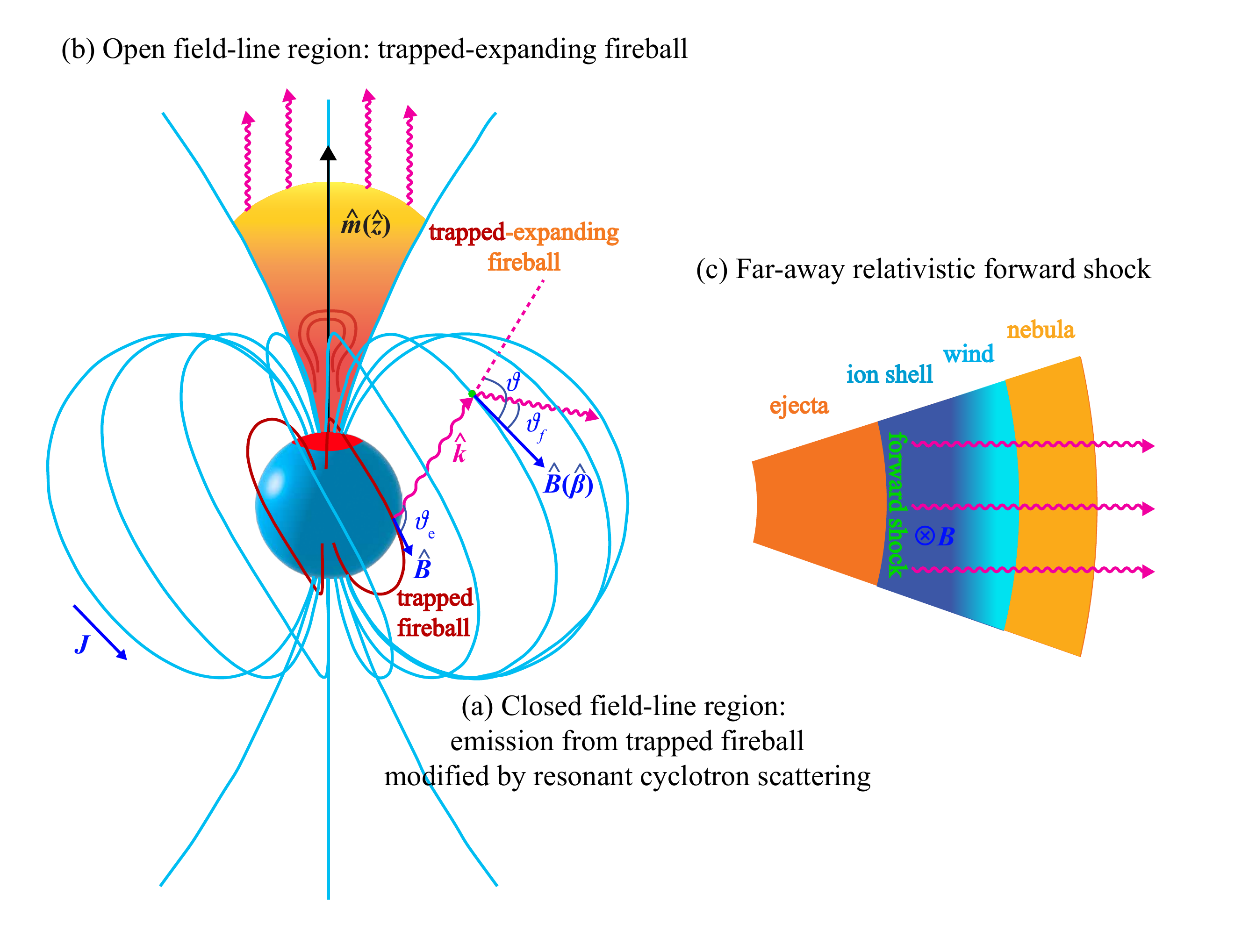}
	\caption{The picture illustrates three commonly-discussed models that generate an FXRB: (a) the emission from a trapped fireball modified by RCS, (b) an outflow from a trapped-expanding fireball along open magnetic field lines, and (c) the far-away synchrotron radiation in a forward shock due to the interaction of relativistic ejecta and ion shell.}
	\label{fig:picture}
\end{figure*}

\section{The Extraordinary FXRB}
\label{sec:extraordinary}
There are two aspects that show the FXRB associated with FRB 20200428D is extraordinary, compared with other ordinary XRBs (OXRBs) from SGR 1935:
(a) The FXRB has a highest peak energy $E_{\rm p}\sim85$ keV in $\nu F_{\nu}$ spectrum among all XRBs from SGR 1935, 
whereas its energy and luminosity are ordinary \citep{rid21}.
(b) The FRB arrival time (i.e., the FXRB arrival time) aligns with
the principal peak of the persistent emission pulse profile \citep[the figure 5 in][]{younes20b}.
Note that the peak is usually presumed to be
the moment when an observer views a hot spot on the neutron star (NS) surface
\citep[e.g.,][]{per08,alba10,younes20a}.
Another notable observation is that the persistent emission flux decreases rapidly with an e-folding time $\sim0.65$ days in early stages of the outburst \citep[the figure 8 in][]{younes20b}, but this observation cannot be another plausible aspect to discriminate the FXRB from OXRBs. This is because the rapid decay should not be directly correlated to the FXRB since it had begun earlier than the FXRB over a dozen hours.

The extraordinary features of the FXRB indicate that the non-detection of other FRB-XRB associations observed by several telescopes \citep{lin20,boch20,kir21} 
could be due to not only the high collimation of FRBs, but also the intrinsic requirements for an FRB-XRB association like the locale suggested in \cite{younes21}.
No matter what, any models trying to interpret an FRB-XRB association should consider the discrepancy between the FXRB and OXRBs.
In general, there are three commonly-discussed models to generate such an FXRB: 
(a) the first one is that the emission from a trapped fireball is modified by resonant cyclotron scattering (RCS) \citep{yam20,yang21}, 
(b) the second is an outflow from a trapped-expanding fireball along open magnetic field lines \citep{ioka20,wada23}, 
and (c) the third is the synchrotron radiation in a forward shock due to the interaction of relativistic flare and ion shell far outside the magnetosphere \citep{met19,mar20}.
These three models are plotted in Figure \ref{fig:picture}. Actually, there is another model to well account for the discrepancy between the FXRB and OXRBs by a quasi-polar/non-polar dichotomy of fireballs owing to photon splitting \citep{younes21}, a magnetar-asteroid impact \citep{dai20}, or a magnetic reconnection event \citep{xie23}. 
However, the photon splitting in the former could be neglected\footnote{This is because for SGR 1935 with a surface magnetic field strength $<10^{15}$ G and the FXRB with $E_{\rm p}\sim85$ keV, the timescale of the photon splitting is likely longer than the dynamical timescale at the radius where photons escape from the fireball \citep{wada23}.} for a trapped-expanding fireball giving rise to the FXRB as suggested by \cite{wada23}, the middle needs an asteroid which will be studied elsewhere, 
while the latter needs a very special local plasma magnetization parameter at the order of $10^3$ \citep{xie23} to explain the first extraordinary feature of the FXRB 
but still cannot explain its second extraordinary feature. Hence we do not consider the photon splitting, asteroid impact, or magnetic reconnection event in this work. 
It should be also emphasized that the third model in Figure \ref{fig:picture} may not be able to offer a good interpretation for the discrepancy between the FXRB and OXRBs. 

To identify the true origin of a similar FXRB, we will study the possible X-ray polarization signatures of each model as follows.

\begin{figure*}
	\centering
	\includegraphics[width=1.0\textwidth]{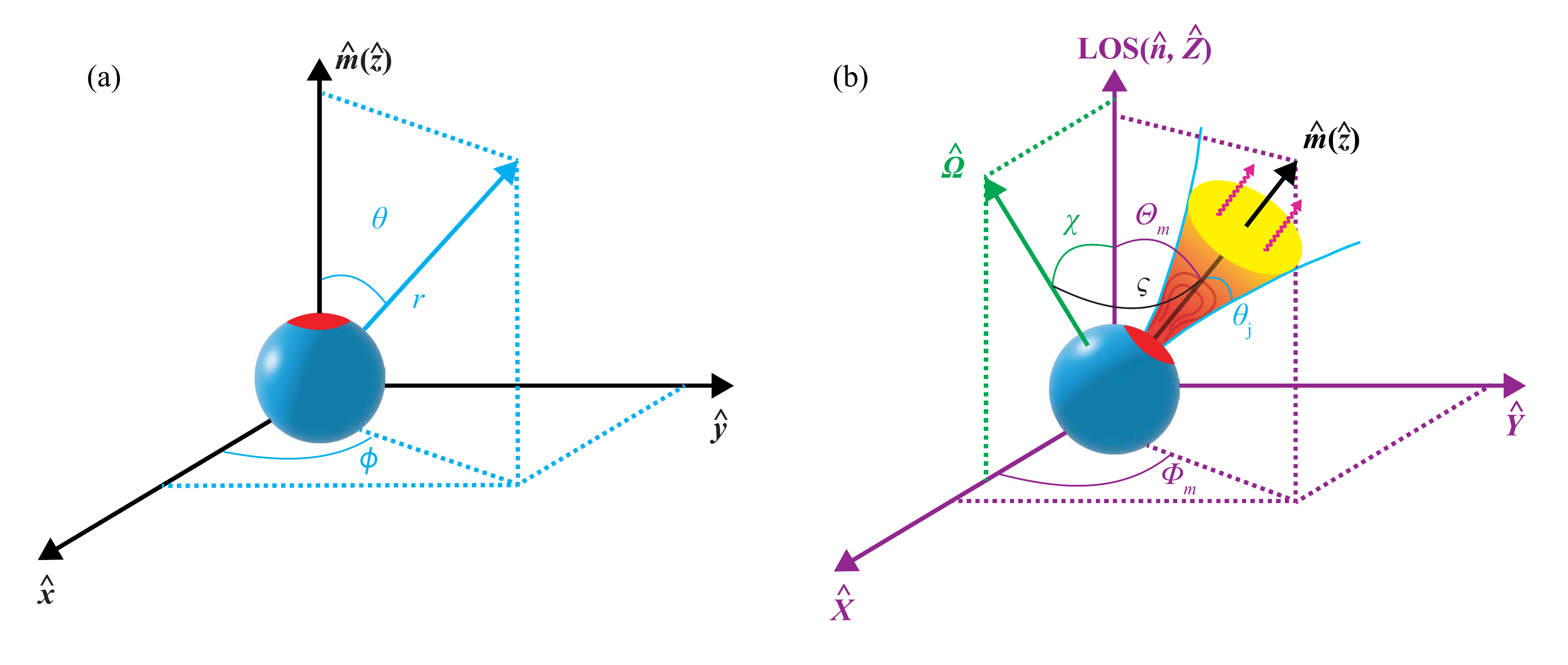}
	\caption{Illustration of two coordinate systems. (a) One is a Cartesian coordinate system ($x$, $y$, $z$) with the $z$-axis along the magnetic axis (unit vector $\boldsymbol{\hat{m}}$), sharing the same origin $\boldsymbol{O}$ with its corresponding spherical coordinate system ($r$, $\theta$, $\phi$). In the latter, $r$ is the radius, $\theta$ and $\phi$ are the magnetic colatitude and azimuth, respectively. (b) The other is a system ($X$, $Y$, $Z$) whose $Z$-axis in the direction of the LOS (unit vector $\boldsymbol{\hat{n}}$), $X$-axis in the plane of $\boldsymbol{\hat{n}}$ and the star spin axis (unit vector $\boldsymbol{\hat{\Omega}}$). Correspondingly, its spherical coordinate system ($r$, $\Theta$, $\Phi$) has inclination $\Theta$ and azimuth $\Phi$ with respect to the LOS. Thus $\boldsymbol{\hat{m}}$ has a coordinate ($\Theta_m$, $\Phi_m$) in this system. The angles that the spin axis makes with the LOS and the magnetic axis denote $\chi$ and $\varsigma$, respectively.}
	\label{fig:coordinates}
\end{figure*}

\section{Coordinate Systems and Transformations}
\label{sec:frame}
For convenience, we firstly introduce two coordinate systems and their transformations before coming into the models.
One is a Cartesian coordinate system ($x$, $y$, $z$) with the $z$-axis along the magnetic axis (unit vector $\boldsymbol{\hat{m}}$), sharing the same origin $\boldsymbol{O}$ (the NS center) with its corresponding spherical coordinate system ($r$, $\theta$, $\phi$). In the latter, $r$ is the radius, $\theta$ and $\phi$ are the magnetic colatitude and azimuth, respectively, see the left panel of Figure \ref{fig:coordinates}. The other is a system ($X$, $Y$, $Z$) whose $Z$-axis in the direction of the light of sight (LOS, unit vector $\boldsymbol{\hat{n}}$), 
$X$-axis in the plane of $\boldsymbol{\hat{n}}$ and the star spin axis (unit vector $\boldsymbol{\hat{\Omega}}$), refer to the right panel of Figure \ref{fig:coordinates}. Correspondingly, its spherical coordinate system ($r$, $\Theta$, $\Phi$) has inclination $\Theta$ and azimuth $\Phi$ with respect to the LOS. All of systems have the same origin.
Moreover, the angles that the spin axis makes with the LOS and the magnetic axis denote $\chi$ and $\varsigma$, respectively. As the star rotates, the angles that the LOS makes with the magnetic axis, denoting $\Theta_m$ and $\Phi_m$ in the system ($r$, $\Theta$, $\Phi$) as the right panel of Figure \ref{fig:coordinates}, change with phase by  
\begin{equation}
	\begin{aligned}
		\cos\Theta_m & = \cos\chi \cos\varsigma + \sin\chi \sin\varsigma \cos\lambda \\
		\cos\Phi_m & = \frac{\cos\chi - \cos\Theta_m \cos\varsigma} {\sin\Theta_m \sin\varsigma},
	\end{aligned}
	\label{eq:rotation}
\end{equation}
where $\lambda$ is the rotation phase. If lacking north–south symmetry as in Section \ref{sec:rcs}, 
there would be $0\leqslant\chi\leqslant\pi$, $0\leqslant\varsigma\leqslant\pi/2$, and $0\leqslant\lambda\leqslant 2\pi$.

\subsection{Magnetic Field}
\label{subsec:field}
Based on Figure \ref{fig:coordinates} and \cite{tave15}, the components of magnetic field $\boldsymbol{B}$ of an NS in the Cartesian coordinate system ($x$, $y$, $z$) 
link to those in the spherical coordinate system ($r$, $\theta$, $\phi$) by
\begin{equation}
	\begin{aligned}
		B_x & =\sin\theta \cos\phi B_r + \cos\theta \cos\phi B_{\theta} - \sin\phi B_{\phi}  \\
		B_y & =\sin\theta \sin\phi B_r + \cos\theta \sin\phi B_{\theta} + \cos\phi B_{\phi} \\
		B_z & =\cos\theta B_r - \sin\theta B_{\theta}.
	\end{aligned}
	\label{eq:B_xyz}
\end{equation}
where 
\begin{equation}
	\left(\begin{array}{c}
		B_r \\
		B_\theta \\
		B_\phi
	\end{array}\right)=\frac{B_{\mathrm{p}}}{2}\left(\frac{R_{\mathrm{NS}}}{r}\right)^3\left(\begin{array}{c}
		2 \cos \theta \\
		\sin \theta \\
		0
	\end{array}\right)
	\label{eq:B_rtp}
\end{equation}
for a pure dipole field structure, where $B_{\rm p}$ is the surface field strength of NS at the pole and $R_{\rm NS}$ is the NS radius. This structure is used in Section \ref{sec:quasi-polar} for the polar trapped-expanding fireball model. While its expression for an axisymmetric, self-similar, globally twisted dipole 
field structure can refer to Equation (\ref{eq:B-vector}), which is used in Section \ref{sec:rcs} for the model invoking an RCS occurring in the closed field-line region.

While its correspondence in the system ($X$, $Y$, $Z$) can be obtained from
\begin{equation}
	\begin{aligned}
		B_X & = B_x \hat{x}_X  + B_y \hat{y}_X + B_z \hat{z}_X \\
		B_Y & = B_x \hat{x}_Y  + B_y \hat{y}_Y + B_z \hat{z}_Y \\
		B_Z & = B_x \hat{x}_Z  + B_y \hat{y}_Z + B_z \hat{z}_Z.
	\end{aligned}
	\label{eq:B_XYZ}
\end{equation}
Above which $\hat{x}_X$, $\hat{x}_Y$, and $\hat{x}_Z$ are the components of the unit vector $\boldsymbol{\hat{x}}$ (i.e., the direction of the $x$-axis of the system ($x$, $y$, $z$)), 
in the system ($X$, $Y$, $Z$), are written as
\begin{equation}
	\begin{aligned}
		\hat{x}_X & = -\sin\chi \sin\varsigma - \cos\chi \cos\varsigma \cos\lambda \\
		\hat{x}_Y & = \cos\varsigma \sin\lambda \\
		\hat{x}_Z & = \sin\chi \cos\varsigma \cos\lambda - \cos\chi \sin\varsigma.
	\end{aligned}
	\label{eq:x_XYZ}
\end{equation}
Similarly, those of the unit vectors $\boldsymbol{\hat{y}}$ and $\boldsymbol{\hat{z}}$ are
\begin{equation}
	\begin{aligned}
		\hat{y}_X & = -\cos\chi \sin\lambda \\
		\hat{y}_Y & = -\cos\lambda \\
		\hat{y}_Z & = \sin\chi \sin\lambda,
	\end{aligned}
	\label{eq:y_XYZ}
\end{equation}
and
\begin{equation}
	\begin{aligned}
		\hat{z}_X & = \sin\chi \cos\varsigma - \cos\chi \sin\varsigma \cos\lambda \\
		\hat{z}_Y & = \sin\varsigma \sin\lambda \\
		\hat{z}_Z & = \cos\chi \cos\varsigma + \sin\chi \sin\varsigma \cos\lambda.
	\end{aligned}
	\label{eq:z_XYZ}
\end{equation}
As a consequence, the angles that the spin axis makes with the LOS ($\chi$) and the magnetic axis ($\varsigma$) are once given, 
the description for the magnetic field in the system ($X$, $Y$, $Z$) will be completely derived at each rotation phase $\lambda$, combining with
\begin{equation}
	\begin{aligned}
		\cos \theta=&\frac{R_{\rm NS}}{r} \sin \Theta \left(\cos \Phi \sin \chi \cos \varsigma+\sin \Phi \sin \varsigma \sin \lambda \right. \\
		& \left.-\cos \Phi \cos \chi \sin \varsigma \cos \lambda \right) \\
		& +\sqrt{1-\left(\frac{R_{\rm NS}}{r} \sin \Theta \right)^2}(\cos \chi \cos \varsigma +\sin \chi \sin \varsigma \cos \lambda),
	\end{aligned}
\end{equation}
and
\begin{equation}
	\begin{aligned}
		\cos \phi=&\frac{R_{\rm NS} \sin \Theta}{r \sin\theta}\left(\sin \Phi \cos \varsigma \sin \lambda-\cos \Phi \sin \chi \sin \varsigma \right. \\
		&\left.-\cos \Phi \cos \chi \cos \varsigma \cos \lambda\right) \\
		& +\sqrt{\frac{r^2-\left(R_{\rm NS} \sin \Theta \right)^2}{r^2 \sin ^2 \theta}}(\sin \chi \cos \varsigma \cos \lambda-\cos \chi \sin \varsigma). 
	\end{aligned}
\end{equation}

\subsection{Ray Trajectory}
\label{subsec:photon}
Within the magnetosphere of an NS, 
a trapped fireball is trapped by the closed magnetic field lines with equation 
$r=R_{\max}\sin^2\theta$ in which $R_{\max}$ is the maximal distance between a given field line to the stellar center.
For a photon emitted from the escaping surface of the trapped fireball, its initial locale $\boldsymbol{r_{\rm e}}=(x_{\rm e}$, $y_{\rm e}$, $z_{\rm e})$, in the Cartesian coordinate system ($x$, $y$, $z$), can be described as
\begin{equation}
	\begin{aligned}
		x_{\rm e} & =r_{\rm e} \sin\theta_{\rm e} \cos\phi_{\rm e}   \\
		y_{\rm e} & =r_{\rm e} \sin\theta_{\rm e} \sin\phi_{\rm e} \\
		z_{\rm e} & =r_{\rm e} \cos\theta_{\rm e},
	\end{aligned}
	\label{eq:xyz_e}
\end{equation}
by transferring that point ($r_{\rm e}$, $\theta_{\rm e}$, $\phi_{\rm e}$) in the spherical coordinate system ($r$, $\theta$, $\phi$), where $r_{\rm e}=R_{\max}\sin^2\theta_{\rm e}$.
The wave vector $\boldsymbol{\hat{k}}$ of this photon with an initial direction ($\vartheta_{\rm e}$, $\varphi_{\rm e}$) referred to the local $\boldsymbol{B}$ (Figure \ref{fig:picture}),  
in the spherical coordinate system,
is given as \citep[cf. the equations (5), (9), and (10) in][]{hu19}
\begin{equation}
	\begin{aligned}
		k_r             & =\frac{2\cos\theta_{\rm e}\cos\vartheta_{\rm e}-\sin\theta_{\rm e}\sin\vartheta_{\rm e}\cos\varphi_{\rm e}}{\sqrt{3\cos^2\theta_{\rm e}+1}}   \\
		k_{\theta} & =\frac{\sin\theta_{\rm e}\cos\vartheta_{\rm e}+2\cos\theta_{\rm e}\sin\vartheta_{\rm e}\cos\varphi_{\rm e}}{\sqrt{3\cos^2\theta_{\rm e}+1}} \\
		k_{\phi}     & =\sin\vartheta_{\rm e}\sin\varphi_{\rm e},
	\end{aligned}
	\label{eq:k_rtp}
\end{equation}
where we have corrected $k_{\theta}$ that has a typo in the second line of the equation (9) in \cite{hu19}.
Such that its correspondence in the Cartesian coordinate system is 
\begin{equation}
	\begin{aligned}
		k_x & = k_r \sin\theta_{\rm e}\cos\phi_{\rm e} + k_{\theta} \cos\theta_{\rm e}\cos\phi_{\rm e} - k_{\phi}\sin\phi_{\rm e} \\
		k_y & = k_r \sin\theta_{\rm e}\sin\phi_{\rm e} + k_{\theta} \cos\theta_{\rm e}\sin\phi_{\rm e} + k_{\phi}\cos\phi_{\rm e} \\
		k_z & = k_r \cos\theta_{\rm e} - k_{\theta} \sin\theta_{\rm e}.
	\end{aligned}
	\label{eq:k_xyz}
\end{equation}
Consequently, the photon trajectory is calculated as
\begin{equation}
	\boldsymbol{r}=\boldsymbol{r_{\rm e}}+ct\boldsymbol{\hat{k}} = 
	\begin{cases}
		x_{\rm e} + ct k_x  \\
		y_{\rm e} + ct k_y  \\
		z_{\rm e} + ct k_z,
	\end{cases}
	\label{eq:xyz}
\end{equation}
where $t$ is the propagation time and $c$ is the speed of light. For simplicity, here we have neglected general relativistic effects around an NS, such as the relativistic ray-bending.

\section{Emission from Trapped Fireball Modified by RCS}
\label{sec:rcs}
Trapped fireball has already been the standard model to account for tail emission of giant flares or
less powerful but more common XRBs \citep{thom95,thom01,kas17}.
For the RCS process, it has been used to upscatter the thermal photons emitted by the cooling surface of a magnetar and fill
the non-thermal tail of the spectrum of quiescent magnetar emission in a
twisted magnetosphere \citep{thom02,bar07,bel07,fer07}.

The X-ray polarization signature in quiescent magnetar emission modified by RCS
has been comprehensively studied by \cite{fer11}, see also \cite{nob08} and \cite{tave14}.
While for magnetar flare emission such as an XRB,
however, its polarization signature may be different from that in magnetar quiescent emission.
This is because:
(a) A different source of seed emission, i.e., flare emission invoking a trapped fireball,
while quiescent emission relevant to an extended region on stellar surface.
(b) A different particle velocity distribution in RCS
due to a large distinction in radiative luminosity with $\sim3-12$ orders of magnitude between flare emission and quiescent emission \citep{bel13,yam20}.

Although the spectral modification of flare emission by RCS with a single scattering in an untwisted dipole magnetic field (i.e., in which charges are accelerated by radiation force due to XRB luminosity as high as $\sim10^{40}~{\rm erg~s^{-1}}$) 
has been explored by \cite{yam20} 
and tried to apply to the FXRB associated with FRB 20200428D, it cannot well reproduce the FXRB spectrum, suggesting a requirement of multiple-scattering RCS
\citep{yam22}. Nonetheless, this model has a challenge to address the issue why only the FXRB has a higher $E_{\rm p}$ by RCS but other XRBs (i.e., OXRBs) also from SGR 1935 do not. Note that the FXRB has a comparable energetics with those OXRBs. The difficulty for this issue in this model may indicate that the charges used for RCS are accelerated by the field twist instead of the radiation force of flare emission.
Therefore, in this section we propose that the FXRB should be related to a field twist that accelerates charges applied to RCS, based on the result that a more twisted magnetic field corresponding to a more significant rise of a power-law tail in spectrum, which has been simulated in quiescent magnetar emission \citep{fer07,nob08}. 
On the other hand, the FRB luminosity is proportional to the magnetic field perturbation in the model of decay of Alfv\'{e}n waves \citep{kumar20}. 
If the field twist needed for FXRB is positively correlated to the magnetic field perturbation needed for FRB or they are intrinsically the same thing, 
the issue can be naturally settled. As this is out of scope of this work, we will pursue it elsewhere.
Anyhow, we will comprehensively study the spectrum and polarization properties of flare emission by containing vacuum polarization and possible multiple-scattering RCS in a twisted magnetosphere below, but ignoring the radiation force that could regulate charge velocity distribution because of the aforementioned reason. 
More specifically, we will perform 3D Monte Carlo (MC) simulations to model the spectral and polarization distributions of seed photons, the propagation of these photons in the magnetosphere affected by RCS, and the final spectrum as well as polarization of outgoing photons, 
based on the methods in \cite{fer07}, \cite{nob08}, \cite{fer11}, and \cite{tave14}
\footnote{We write the numerical simulation code mainly based on the public primary and incomplete code written by Dr. Andrei Igoshev in {\em https://github.com/ignotur/magnetar\_spectrum}.}.

\subsection{RCS}
\label{subsec:scattering}
We consider a scenario in which the photons are emitted from a trapped fireball and then scattered by
the electrons in a twisted magnetosphere, as shown in Figure \ref{fig:picture}.
In this scenario, an incident photon with angular frequency $\omega$ measured in the stellar frame\footnote{For further detail, one can see, e.g., the equation (2) in \cite{yam20}.} 
(SF; i.e., lab frame) will be resonantly scattered
by an electron with velocity $v=\beta c$ (Lorentz factor $\gamma={1/\sqrt{1-\beta^2}}$) relative to the SF when the condition
\begin{equation}
	\omega=\omega_D\equiv\frac{\omega_B}{\gamma\left(1-\beta \mu \right)}
	\label{eq:omega}
\end{equation}
is satisfied, where $\omega_B=eB/m_e c$ is the cyclotron frequency and
$\mu=\cos \vartheta$ is the cosine of the incident angle between the photon direction $\boldsymbol{\hat{k}}$ and the electron
velocity direction $\boldsymbol{\hat{\beta}}$ (i.e., the magnetic field direction $\boldsymbol{\hat{B}}$) all measured in the SF, see Figure \ref{fig:picture}.
Because the magnetic field $\boldsymbol{B}$ is not affected by the Lorentz transformation as electrons are moving along the field lines,
the cyclotron frequency $\omega_B$ in the SF is the same as $\omega_B^{\prime}$ in the electron rest frame (ERF).
In Section \ref{sec:rcs}, the quantities without a prime are in the SF, while those with a prime are in the ERF if not otherwise specified.
For a RCS process, the spectral and polarization distributions of incident (seed) photons, 
the charge spatial and velocity distributions, the scattering cross-sections, as well as the scattering photons must be clear. 

\subsubsection{Polarization, Energy, and Spatial Distributions of Seed Photons}
\label{subsubsec:seed}
The vacuum around star with a strong magnetic field behaves as a birefringent medium due to vacuum polarization.
If vacuum polarization dominates in a trapped fireball,
photons within the trapped fireball propagate outward in two normal modes of polarization:
the ordinary mode (O-mode) with the electric vector in the plane of the wave vector $\boldsymbol{\hat{k}}$
and the background magnetic field $\boldsymbol{\hat{B}}$, and the extraordinary mode
(E-mode) with the electric vector perpendicular to this plane \citep[e.g.,][]{mes92,hard06}.
When the primary photon energy $\varepsilon=\hbar \omega \ll \hbar \omega_B$ measured in the SF,
the escaping seed photons are mostly polarized in the E-mode, since the cross-section of E-mode photons is much less than that for O-mode photons, i.e., $\sigma_{\rm E}/\sigma_{\rm O}\sim\left(\varepsilon / \hbar\omega_{B}\right)^{2}\simeq 10^{-4}(\varepsilon / 10 {\rm keV})^2\left(B / 10^{14}{\rm G}\right)^{-2}$ \citep[e.g.,][]{mes92,hard06}. 
For the FXRB associated with FRB 20200428D, $\sigma_{\rm E}/\sigma_{\rm O}<0.02$ for its spectral energy range $\sim10-300$ keV \citep{rid21},
under the condition of the surface magnetic field strength at the pole of SGR 1935 $B_{\rm p}=2\times10^{14}~{\rm G}$ \citep{isr16}.

As a result, one may have the energy distribution of the seed photons as a non-Planckian form from the integration over each layer of E-mode photosphere \citep{lyu02}
\begin{equation}
	N(\varepsilon)=0.47 \varepsilon^{2}\left\{\exp \left[\frac{\varepsilon^{2}}{T_{\rm b} \sqrt{\varepsilon^{2}+\left(3 \pi^{2} / 5\right) T_{\rm b}^{2}}}\right]-1\right\}^{-1},
	\label{eq:N-epsilon}
\end{equation}
where $T_{\rm b}$ is the bolometric temperature of the trapped fireball with the same energy unit as $\varepsilon$, e.g., keV.
In this section we use this simple description for the energy distribution of the seed photons and 
randomly generate each seed photon energy $\varepsilon$ from this distribution and set its polarization as E-mode, 
even though the polarization of more realistic emission from a
trapped fireball is likely geometric-structure dependent \citep{yang15,tave17}.

Regarding the spatial distribution of the seed photons, one can start from the constraints on the trapped fireball. 
As it is known that the FXRB has a total radiated energy $E_{\rm FXRB}\sim10^{40}~{\rm erg}$ and luminosity $L_{\rm FXRB}\sim10^{41}~{\rm erg~s^{-1}}$
\citep{mere20,lick21,rid21,tava21}. If the
energy is trapped by the close field lines of a purely dipole magnetic field, one may have the scale upper limit of the trapped fireball above 
the stellar surface $\Delta R\lesssim80R_{\rm NS}$ from
$\frac{B_{R_{\rm NS}+\Delta R}^{2}}{8 \pi}\sim  \frac{B_{\rm p}^2  (\frac{R_{\rm NS}+\Delta R}{R_{\rm NS}})^{-6}}{8 \pi} \gtrsim \frac{E_{\rm FXRB}}{3 \Delta R^{3}}$ 
\citep[see the inequation (1) in][]{thom95} for $R_{\rm NS}=10^6~{\rm cm}$.
On the other hand, if the FXRB is generated by RCS from an OXRB with a typical temperature $T\sim10T_1$ keV (here $T_1=T/10^1{\rm keV}$), the size of the trapped fireball corresponding to the OXRB is estimated as $\Delta R = \left(\frac{L_{\rm FXRB}}{2\pi c a T^4}\right)^{1/2}\sim 6\times10^5~{\rm cm}~L_{\rm FXRB,41}^{1/2}T_1^{-2}$ in which $a$ is the radiation constant \cite[e.g.,][]{ioka20}.
Then the border of the closed field lines enclosing the trapped fireball can be described by the field line equation $r=R_{\max}\sin^2\theta$ 
in which $R_{\max}\sim R_{\rm NS}+\Delta R \sim 2R_{\rm NS}$.
Note that the field line equation is only dependent on $\theta$ for a purely dipole magnetic field, but it is also dependent on the azimuthal angle $\phi$ for a twisted magnetic field. For convenience, we adopt the field line equation $r=R_{\max}\sin^2\theta$ that has no $\phi$ dependence since it is a good approximation for a moderately twisted magnetic field.
Given that the radiation flux is the same everywhere on the escaping surface of the trapped fireball, one initial photon should emit from a locale ($r_{\rm e}=R_{\max}\sin^2\theta_{\rm e}$, $\theta_{\rm e}$, $\phi_{\rm e}$) with $\theta_{\rm e}$ and $\phi_{\rm e}$ yielded respectively from uniform random numbers $U_1=\int_{\theta_{\rm e,min}}^{\theta_{\rm e}}dS/\int_{\theta_{\rm e,min}}^{\theta_{\rm e,max}}dS\sim U(0, 1)$\footnote{where $dS=2\pi R_{\max}^2\sin^4\theta\sqrt{1+3\cos^2\theta}d\theta$ is the differential area of the closed field lines with $R_{\max}$ \citep{yang15}, ${\theta_{\rm e,min}}=\arcsin\left(\sqrt{R_{\rm NS}/R_{\max}}\right)$ and ${\theta_{\rm e,max}}=\pi-{\theta_{\rm e,min}}$ \citep{tave17}.} 
and $U_2\sim U(0, 2\pi)$. 
Furthermore, under the assumption that initial direction of the emitted photon is isotropic, 
one can draw the initial direction by 
the angle $\vartheta_{\rm e}$ (the photon direction $\boldsymbol{\hat{k}}$ making with the local field direction $\boldsymbol{\hat{B}}$)
and the azimuthal angle $\varphi_{\rm e}$ as referred to $\boldsymbol{\hat{B}}$ at the emission locale (see Figure \ref{fig:picture}), 
which can be yielded from uniform random numbers $U_3\sim U(0, \pi)$ and $U_4\sim U(0, 2\pi)$, respectively. 
It is convenient to take the zero of $\varphi_{\rm e}$ to coincide with the $\boldsymbol{\hat{B}}$-$\boldsymbol{\hat{z}}$ plane, where $\boldsymbol{\hat{z}}$ is along the magnetic axis $\boldsymbol{\hat{m}}$.

\subsubsection{Charge Spatial and Velocity Distributions in a Twisted Magnetosphere}
\label{subsubsec:charge}
An axisymmetric, self-similar, globally twisted dipole magnetosphere is given by \citep{thom02}
\begin{equation}
	\boldsymbol{B}=\frac{B_{\mathrm{p}}}{2}\left(\frac{r}{R_{\mathrm{NS}}}\right)^{-p-2}\left[F_r, F_{\theta}, F_{\phi}\right],
	\label{eq:B-vector}
\end{equation}
where $F_r=-f^{\prime}$ in which a prime denotes derivation with respect to $\cos\theta$, $F_{\theta}=\frac{p f}{\sin \theta}$, and $F_{\phi}=\sqrt{\frac{C(p) p}{p+1}} \frac{f^{1+1/p}}{\sin \theta}$.
The function $f=f(\cos\theta)$ is the solution to the Grad–Shafranov equation
It satisfies three boundary conditions $f^{\prime}(0)=0$, $f^{\prime}(1)=-2$, and $f(1)=0$.
Both $f$ and $C$ can be numerically calculated once the value of $p$ relevant to the field twist is fixed.
Besides $p$, the amount of the twist is generally measured by the twist angle 
\begin{equation}
	\begin{aligned}
		\Delta \phi_{\mathrm{N}-\mathrm{S}} & =2\lim _{\theta \rightarrow 0} \int_\theta^{\pi / 2} \frac{B_\phi}{B_\theta \sin \theta} \mathrm{d} \theta \\
		& =2\left[\frac{C(p)}{p(1+p)}\right]^{1 / 2} \lim _{\theta \rightarrow 0} \int_\theta^{\pi / 2} \frac{f^{1 / p}}{\sin \theta} \mathrm{d} \theta .
	\end{aligned}
	\label{eq:Delta-phi}
\end{equation}
where $\Delta \phi_{\mathrm{N}-\mathrm{S}}$ ranges from 0 to $\pi$ ($p$ from 1 to 0).

The spatial density distribution of the charges along field lines in this twisted magnetosphere is then
\begin{equation}
	n_e(\boldsymbol{r},\beta)=\frac{p+1}{4 \pi e}\left(\frac{B_\phi}{B_\theta}\right) \frac{B}{r|\langle\beta\rangle|},
	\label{eq:n_e}
\end{equation}
where $\langle\beta\rangle$ is the average charge velocity in units of $c$ \citep{nob08}.
For the simplest case, the charges are assumed to be the unidirectional flow electrons moving from the north to the south pole and have a 1D relativistic Maxwellian distribution at a given temperature $T_e$ superimposed to a bulk motion with velocity $\beta_{\rm b}$ measured in the SF.
Such that the velocity (momentum $\gamma\beta$) distribution is given by
\begin{equation}
	\frac{d n_e}{d(\gamma \beta)}=\frac{n_e \exp \left(-\gamma^{\prime} / \Theta_e\right)}{2 K_1\left(1 / \Theta_e\right)}=n_e f_e(\boldsymbol{r}, \gamma\beta),
	\label{eq:electron_velocity}
\end{equation}
where $\gamma^{\prime}=\gamma \gamma_{\text {b}}\left(1-\beta \beta_{\text {b}}\right)$ and $\gamma_{\rm b}=1/\sqrt{1-\beta_{\rm b}^2}$, $\Theta_e=k_{\rm B} T_e / m_e c^2$, 
$K_1$ is the modified Bessel Function of the second kind, 
and $f_e(\boldsymbol{r}, \gamma\beta)=\gamma^{-3}n_e^{-1}dn_e/d\beta$ is the momentum distribution function \citep{nob08}.
If $T_e$ and $\beta_{\rm b}$ regarded as two free parameters are assumed to be both independent of position, the velocity distribution function would be \citep[cf. the equation (23) in][]{fer07}, 
no longer relying on position $\boldsymbol{r}$,
\begin{equation}
	f_e(\beta)=n_e^{-1}\frac{dn_e}{d\beta}=\gamma^3 \frac{\exp \left(-\gamma^{\prime} / \Theta_e\right)}{2 K_1\left(1 / \Theta_e\right)},
	\label{eq:f_e}
\end{equation}
and which is normalized by ($\beta>0$)
\begin{equation}
	\int_{0}^{1} f_e(\beta) d\beta=1.
\end{equation}
So that the average charge velocity can be computed as
\begin{equation}
	\langle\beta\rangle=\int_{0}^{1} \beta f_e(\beta) d\beta.
	\label{eq:beta_ave}
\end{equation}

\subsubsection{Photon Propagation in the Magnetosphere}
\label{subsubsec:propagation}
Due to RCS, an O or E-mode photon released from the trapped fireball which travels a distance $d\ell$ will see a differential optical depth \citep{fer07,nob08}
\begin{equation}
	\begin{aligned}
		d \tau_{\rm O}&=d \tau_{\rm O-O}+d \tau_{\rm O-E}  \\
		&=2 \pi^2 r_0 c \frac{n_e \omega_B}{\omega^2} d \ell \sum_{k=+,-} \frac{\left|\mu-\beta_k\right|}{\left(1-\mu \beta_k\right)} f_e\left(\boldsymbol{r}, \gamma_k \beta_k\right),
	\end{aligned}
	\label{eq:dtau_O}
\end{equation}
or 
\begin{equation}
	\begin{aligned}
		d \tau_{\rm E}&=d \tau_{\rm E-E}+d \tau_{\rm E-O} \\
		&=2 \pi^2 r_0 c \frac{n_e \omega_B}{\omega^2} d \ell \sum_{k=+,-} \frac{\left(1-\mu \beta_k\right)}{\left|\mu-\beta_k\right|} f_e\left(\boldsymbol{r}, \gamma_k \beta_k\right),
	\end{aligned}
	\label{eq:dtau_E}
\end{equation}
respectively, where the first subscript refers to the incident photon polarization mode and the second to the scattered photon, $r_0$ is the classical electron radius, and $\beta_k$ are the roots of the resonance condition in Equation (\ref{eq:omega}) written as
\begin{equation}
	\beta_{\pm}=\frac{1}{\mu^2+\left(\omega_B / \omega\right)^2}\left[\mu \pm \frac{\omega_B}{\omega} \sqrt{\left(\omega_B / \omega\right)^2+\mu^2-1}\right].
	\label{eq:beta_pm}
\end{equation}

The optical depth along the propagation path when the photon travels distance $\ell$ is given by stepwise integrating Equations (\ref{eq:dtau_O}) and (\ref{eq:dtau_E})
\begin{equation}
	\tau_s=\int_0^{\ell} d \tau_s=-\ln U_0,
	\label{eq:tau_s}
\end{equation}
where $s=$ O or E mode and $U_0\sim U(0,1)$ is a uniform random number. When $\tau_s>-\ln U_0$, integration is terminated, the scattering occurs, $\tau_s$ returns back to zero, and a new path of the photon will be run. The new polarization, propagation direction, and energy of the photon after scattering are determined below.

\subsubsection{New Polarization, Direction, and Energy of Photons after Scattering}
\label{subsubsec:afterscattering}
To obtain the new photon polarization, direction, and energy, we follow \cite{nob08} to setup the implementation after scattering:
(1) A uniform random number $U_5\sim U(0,1)$ is generated in order to decide the polarization mode switching upon scattering.
For an incident O-mode (or E-mode) photon, the mode switching occurs when $U_5>\sigma_{\rm O-O}/(\sigma_{\rm O-O}+\sigma_{\rm O-E})=1/4$ (or $U_5>\sigma_{\rm E-E}/(\sigma_{\rm E-E}+\sigma_{\rm E-O})=3/4$).
(2) A uniform random number $U_6\sim U(0,1)$ is generated to decide the velocity of the scattering electron.
If $U_6<S_s\left(\beta_{+}\right) /\left[S_s\left(\beta_{+}\right)+S_s\left(\beta_{-}\right)\right]$ in which $S_s(\beta_k)$ represents each addendum in the sum at left-hand sides of Equations (\ref{eq:dtau_O}) and (\ref{eq:dtau_E}), the electron velocity is $\beta_{+}$, otherwise it is $\beta_{-}$.
(3) Another uniform random number $U_7\sim U(0,1)$ is generated to
decide the azimuthal angle about the local magnetic field direction $\varphi_f=2\pi U_7$ in the SF where the zero of $\varphi_f$ coinciding with the $\boldsymbol{\hat{B}}$-$\boldsymbol{\hat{z}}$ plane. 
(4) A final uniform random number $U_8\sim U(0,1)$ is generated to decide the scattering angle 
$\vartheta^{\prime}_f$ between the photon direction and the local magnetic field direction in the ERF given by $\cos\vartheta^{\prime}_f=2U_8-1$ (for O$-$E or E$-$E mode switching) or $\cos^3\vartheta^{\prime}_f=2U_8-1$ 
(for O$-$O or E$-$O mode switching)\footnote{When $\cos^3\vartheta^{\prime}_f=2U_8-1<0$, $\cos\vartheta^{\prime}_f=-(-\cos^3\vartheta^{\prime}_f)^{1/3}$.}.
Correspondingly, the cosine of the scattering angle in the SF is obtained by, through Lorentz transformation,
\begin{equation}
	\mu_f=\cos\vartheta_f=\frac{\cos\vartheta^{\prime}_f+\beta_k}{1+\beta_k\cos\vartheta^{\prime}_f}.
	\label{eq:mu}
\end{equation}

In addition, the cosine of the magnetic colatitude $\mu_k$ of the scattered photon direction $\hat{\boldsymbol{k}}$ can be read as \citep{fer07}
\begin{equation}
	\mu_k=\mu_B \mu_f+\sqrt{\left(1-\mu_B^2\right)\left(1-\mu_f^2\right)} \cos \varphi_f,
	\label{eq:mu_k}
\end{equation}
where $\mu_B =\hat{\boldsymbol{B}} \cdot \hat{\boldsymbol{z}}$.
Its azimuth $\phi_k$ about the magnetic axis also is calculated by \citep{fer07}
\begin{equation}
	\phi_k=\arctan\left(k_y/k_x\right),
	\label{eq:phi_k}
\end{equation}
where $k_x$ and $k_y$ are the components of the scattered photon direction $\hat{\boldsymbol{k}}$ along $\hat{\boldsymbol{x}}$ and $\hat{\boldsymbol{y}}$ axis, 
which are obtained from
\begin{equation}
	k_z=\mu_k,
\end{equation}
\begin{equation}
	\hat{\boldsymbol{k}} \cdot \hat{\boldsymbol{B}}=\mu_f,
\end{equation}
and 
\begin{equation}
	\hat{\boldsymbol{k}} \cdot(\hat{\boldsymbol{B}} \times \hat{\boldsymbol{z}})=\sqrt{\left(1-\mu_B^2\right)\left(1-\mu_f^2\right)} \sin \varphi_f.
\end{equation}

Through Lorentz transformation again, the scattered photon frequency in the SF is got as \citep{nob08}, 
\begin{equation}
	\omega_f=\gamma_k^2 \omega\left(1-\beta_k \mu\right)\left(1+\beta_k \cos \vartheta^{\prime}_f\right).
	\label{eq:omega_f}
\end{equation}
It is worth stressing that $\omega$ is still the incident photon frequency and 
$\mu$ is still the cosine of the incident angle of the photon direction with respect to the local magnetic field direction before next scattering.

\subsubsection{Outgoing Photons}
\label{subsubsec:outgoing}
In each stepwise integration of Equation (\ref{eq:tau_s}) for each seed photon, we first check whether the position coordinate ($r_p$, $\theta_p$, $\phi_p$) of the photon satisfies $r_p<R_{\max}\sin^2\theta_p$. If it does, the photon backwards to the trapped fireball and we would discard it. If the photon does not backward to the trapped fireball, we further follow the section 3.2 of \cite{nob08} to check whether the photon satisfies the escaping conditions. 
If it does, the photon is taken to freely escape.
Its polarization $s$, direction $\mu_k$ and $\phi_k$, and energy $\varepsilon=\hbar \omega$ are then stored 
\footnote{In this subsection, $\mu$ and $\omega$ are respectively the final incident angle and frequency of the photon despite the photon experiences either multiple scatterings or zero scatterings.}.
For the statistical polarization properties of outgoing photons, we proceed the study below.

\subsection{Polarization Evolution of Photons Propagating in the Magnetosphere}
\label{subsec:polarization}
The polarization state of a photon propagating in the magnetosphere is affected by the vacuum polarization in a strong magnetic field \citep[e.g.,][]{mes92,hard06}.
When taking a reference frame ($x_i$, $y_i$, $z_i$) with the $z_i$-axis along the photon wave vector $\boldsymbol{\hat{k}}$ and the local background magnetic field initially lies in the $x_i-z_i$ plane, the electric vector of the outgoing photon with energy $\hbar\omega$ can be expressed as
\begin{equation}
	\boldsymbol{E}=\boldsymbol{E}_0(z_i) {\rm e}^{-{\rm i} \omega t}=\boldsymbol{A}(z_i) {\rm e}^{{\rm i}\left(k_0 z_i-\omega t\right)},
	\label{eq:E-vector}
\end{equation} 
where $k_0=\omega/c$ and $\boldsymbol{A}(z_i)$ is the complex amplitude.
The amplitude evolution along the traveling path is govern by
\begin{equation}
	\begin{aligned}
		\frac{d A_{x_i}}{d z_i} & =\frac{{\rm i} k_0 \delta}{2}\left[M A_{x_i}+P A_{y_i}\right] \\
		\frac{d A_{y_i}}{d z_i} & =\frac{{\rm i} k_0 \delta}{2}\left[P A_{x_i}+N A_{y_i}\right],
	\end{aligned}
	\label{eq:dA}
\end{equation}
where $\delta =\frac{\alpha_{\rm F}}{45 \pi}\left(\frac{B}{B_{\rm Q}}\right)^2 \simeq 3 \times 10^{-10}\left(\frac{B}{10^{11} {\rm G}}\right)^2$ in which $\alpha_{\rm F}\simeq 1/137$ is the fine-structure constant, $M$, $N$, and $P$ can refer to \cite{tave14}.
Its Stokes parameter form is \citep{tave14}
\begin{equation}
	\begin{aligned}
		& \frac{dQ_i}{dz_i^{\prime}}=-2 P V_i \\
		& \frac{dU_i}{dz_i^{\prime}}=-(N-M) V_i \\
		& \frac{dV_i}{dz_i^{\prime}}=2 P Q_i+(N-M) U_i,
	\end{aligned}
	\label{eq:dQUV}
\end{equation}
where $dz_i^{\prime}=k_0 \delta dz_i/2$. This Stokes parameter form is in a new fixed reference frame ($u$, $v$, $n$) with the $n$-axis along the LOS (i.e., the photon wave vector $\boldsymbol{\hat{k}}$ since only those photons with $\boldsymbol{\hat{k}}$ along the LOS can be observed) and the $u$-axis as well as $v$-axis in the polarimeter plane. 
The Stokes parameters $I_i$, $Q_i$, $U_i$, and $V_i$ in the frame ($u$, $v$, $n$) relate to those with a bar in the frame ($x_i$, $y_i$, $z_i$) through
\begin{equation}
	\begin{aligned}
		I_i & =\bar{I}_i \\
		Q_i & =\bar{Q}_i \cos (2 \alpha_i)+\bar{U}_i \sin (2 \alpha_i) \\
		U_i & =\bar{U}_i \cos (2 \alpha_i)-\bar{Q}_i \sin (2 \alpha_i) \\
		V_i & =\bar{V}_i,
	\end{aligned}
	\label{eq:IQUV}
\end{equation}
where $\alpha_i$ can be obtained by 
\begin{equation}
	\cos\alpha_i = \hat{\boldsymbol{u}} \cdot \hat{\boldsymbol{x}}_i = \frac{B_X \sin \psi-B_Y \cos \psi}{\sqrt{B_X^2+B_Y^2}},
	\label{eq:cos_alpha_i}
\end{equation}
in which $B_X$ and $B_Y$ are in Equation (\ref{eq:B_XYZ}) and are functions of $\chi$, $\varsigma$, and rotation phase $\lambda$ \citep{tave15}. When the angle $\psi=0$ between the $u$-axis and the $X$-axis is adopted, the $u$-axis coincides with the $X$-axis that is in the plane of the LOS and the star spin axis. Furthermore, the determination in the sign of $\alpha_i$ can be solved as $\sin\alpha_i=-\sqrt{1-\cos^2\alpha_i}$ if $B_X>0$ else $\sin\alpha_i=\sqrt{1-\cos^2\alpha_i}$ \citep{tave15}.
Moreover, the initial conditions are set as $\bar{U}_i=\bar{V}_i=0$, $\bar{I}_i=1$, 
and $\bar{Q}_i=\pm1$ in which plus (minus) denotes an O-mode (E-mode) photon. 
When the escaping condition of the photon is met (Section \ref{subsubsec:outgoing}), the integration for the polarization evolution in Equations (\ref{eq:dQUV}) is executed
until the radial distance reaching the adiabatic radius\footnote{Before the adiabatic radius, normal modes do not mix and the propagation is adiabatic \citep{hey00,lai03}.} \citep{fer11}
\begin{equation}
	\begin{aligned}
		r_{\rm a}=R_{\rm NS}\left[\left(\frac{\alpha_{\rm F}}{30 \pi}\right)\left(\frac{B_{\rm p}}{B_{\rm Q}}\right)^2\left(1-\mu^2\right) \frac{\xi^2}{\zeta}\left(\frac{R_{\rm NS} \omega}{c}\right)\right]^{1 /(3+2p)},
	\end{aligned}
	\label{eq:r_a}
\end{equation}
where $\xi=1/2\left(F_r^2+F_{\theta}^2+F_{\phi}^2\right)^{1/2}$ in which $F_r$, $F_{\theta}$, and $F_{\phi}$ are in Equation (\ref{eq:B-vector}) 
and $\zeta$ is a dimensionless function of order unity.

However, the integration for Equations (\ref{eq:dQUV}) may be unnecessary since only the magnetic field direction at the adiabatic radius influences the outgoing electric vector of each photon \citep{hey00,hey02,lai03}. Such that $\alpha_i$ in Equations (\ref{eq:IQUV}) only at the adiabatic radius is required and the Stokes parameters $I_i$, $U_i$, and $Q_i$ for each outgoing photon are then stored along its polarization mode $s$, direction $\theta_k$ as well as $\phi_k$, and energy $\varepsilon$
as mentioned in Section \ref{subsubsec:outgoing}. Finally, an 8D array for each photon are listed in a table, when the initial energy of the photon is also included.  

Accordingly, the Stokes parameters of the outgoing radiation collected on a specific sky patch at infinity ($\theta_k$, $\phi_k$)\footnote{The sky patches at infinity are characterized by the magnetic colatitude $\theta_k$ and azimuth $\phi_k$, similar to the patches on the NS surface.} are, by summing the photons with the same direction $\theta_k$ and $\phi_k$ (and sorting them according to the energy if needed), 
\begin{equation}
	\begin{aligned}
		I & =\sum_{i=1}^N I_i=N \\
		Q & =\sum_{i=1}^N Q_i \\
		U & =\sum_{i=1}^N U_i.
	\end{aligned}
	\label{eq:QUI}
\end{equation}
The linear polarization degree (PD) and polarization angle (PA) are then computed as
\begin{equation}
	\begin{aligned}
		\Pi& =\frac{\sqrt{Q^2+U^2}}{I} \\
		\chi_{\rm p} & =\frac{1}{2} \arctan \left(\frac{U}{Q}\right),
	\end{aligned}
	\label{eq:PD-PA}
\end{equation}
respectively\footnote{If O-mode photons dominate the summing at a specific sky patch, the PA is $\chi_{\rm p}$, else it is $\chi_{\rm p}+\frac{\pi}{2}$. Here we only consider the linear polarization since the circular polarization is not expected to be detectable with forthcoming X-ray instruments, as stated in \cite{tave14}.}.

\begin{deluxetable}{lllll}
	\label{tab:bestfit}
	\tablecaption{One Group of Good Parameter Values for Fitting the Spectrum of the FXRB associated with FRB 20200428D}
	\tablehead{
		\colhead{Parameters} &
		\colhead{Values}
	}
	\startdata
	\object{\bf Trapped Fireball}  &    \\
	\hline
	\object{Maximal radius $R_{\max}$}  &  $2R_{\rm NS}$   \\
	\object{Bolometric temperature $T_{\rm b}$} &  15 keV    \\
	\hline
	\object{\bf Twisted Magnetosphere}     &     \\
	\hline
	\object{Surface magnetic field strength $B_{\rm p}$}     &   $2\times10^{14}$ G   \\
	\object{Twist angle $\Delta\phi_{\rm N-S}$}     &    1 rad   \\
	\object{Charge bulk velocity $\beta_{\rm b}$} &  $0.3c$      \\
	\object{Charge temperature $T_e$}      &  30 keV     \\
	\object{Magnetic colatitude $\theta_k$}      &  $40^{\circ}$     \\
	\hline
	\object{\bf Goodness of Fitting}     &     \\
	\hline
	\object{$\chi^2$/dof}      &  23/27
	\enddata
\end{deluxetable}

\begin{figure*}
	\centering
	\includegraphics[width=8.5cm]{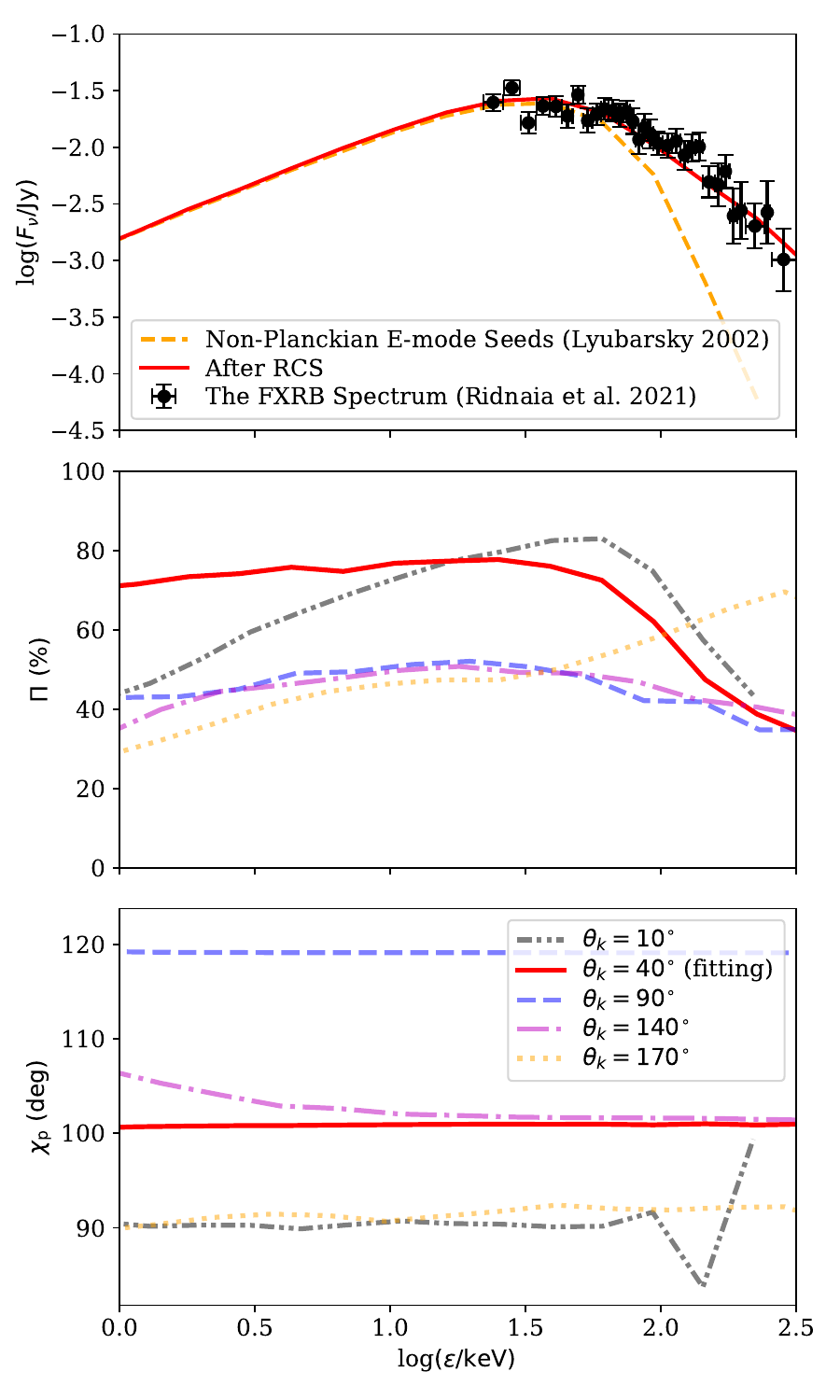}
	\includegraphics[width=8.5cm]{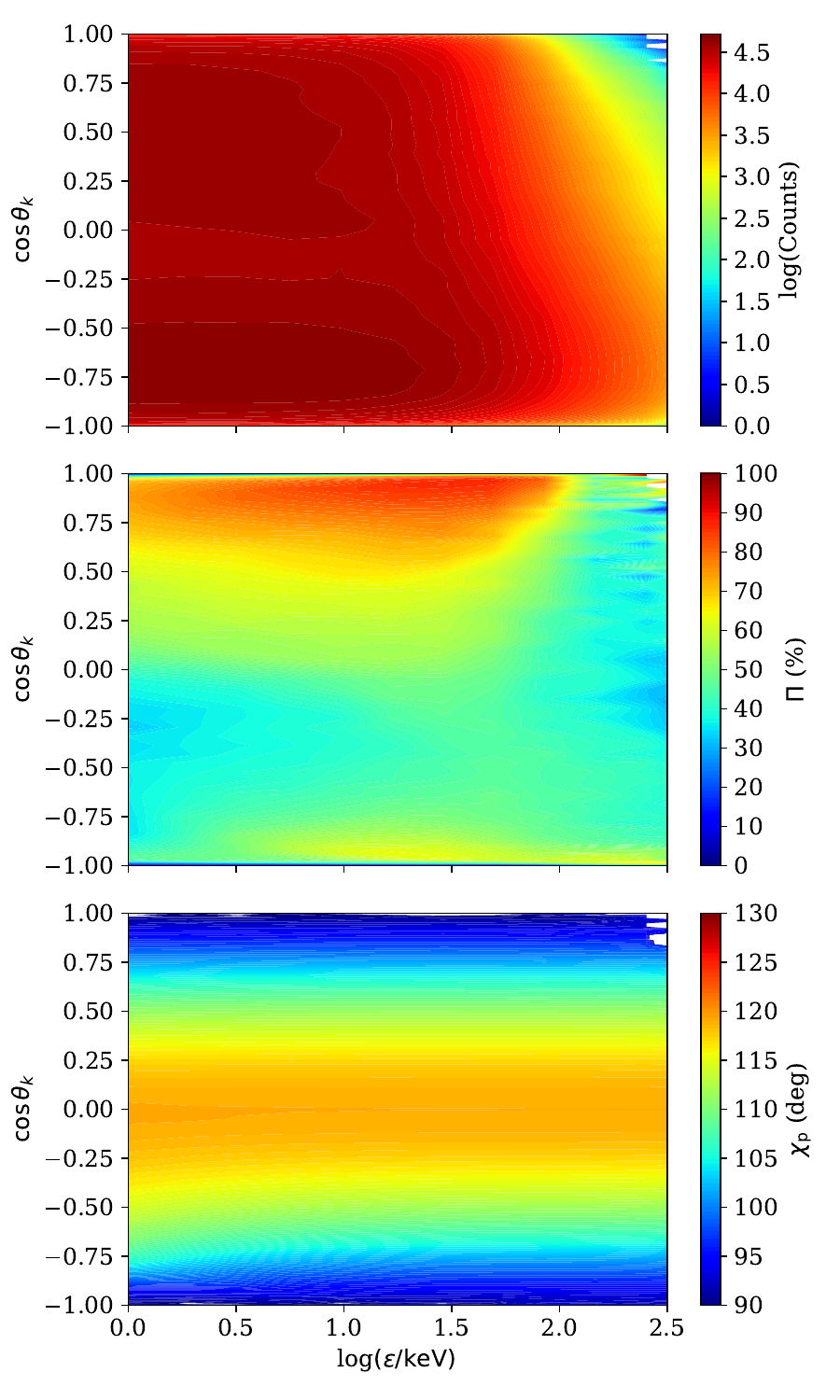}
	\caption{Phase-averaged spectra and polarization properties within the model of the emission from a trapped fireball modified by RCS. {\em Left:}
		{\it Top}: One good fitting for the spectrum of the FXRB associated with FRB 20200428D.
		{\it Middle}: PDs ($\Pi$) as a function of energy ($\varepsilon$) in different values of magnetic colatitude ($\theta_k$), in which
		$\theta_k = 40^{\circ}$ is the value that corresponds to the good fitting for the FXRB spectrum.
		{\it Bottom}: PAs ($\chi_{\rm p}$) as a function of energy in different magnetic colatitudes. 
		{\em Right:} same as left panels, but with contour plots.}
	\label{fig:fig2_spec}
\end{figure*}

\begin{figure*}
	\centering
	\includegraphics[width=5.5cm]{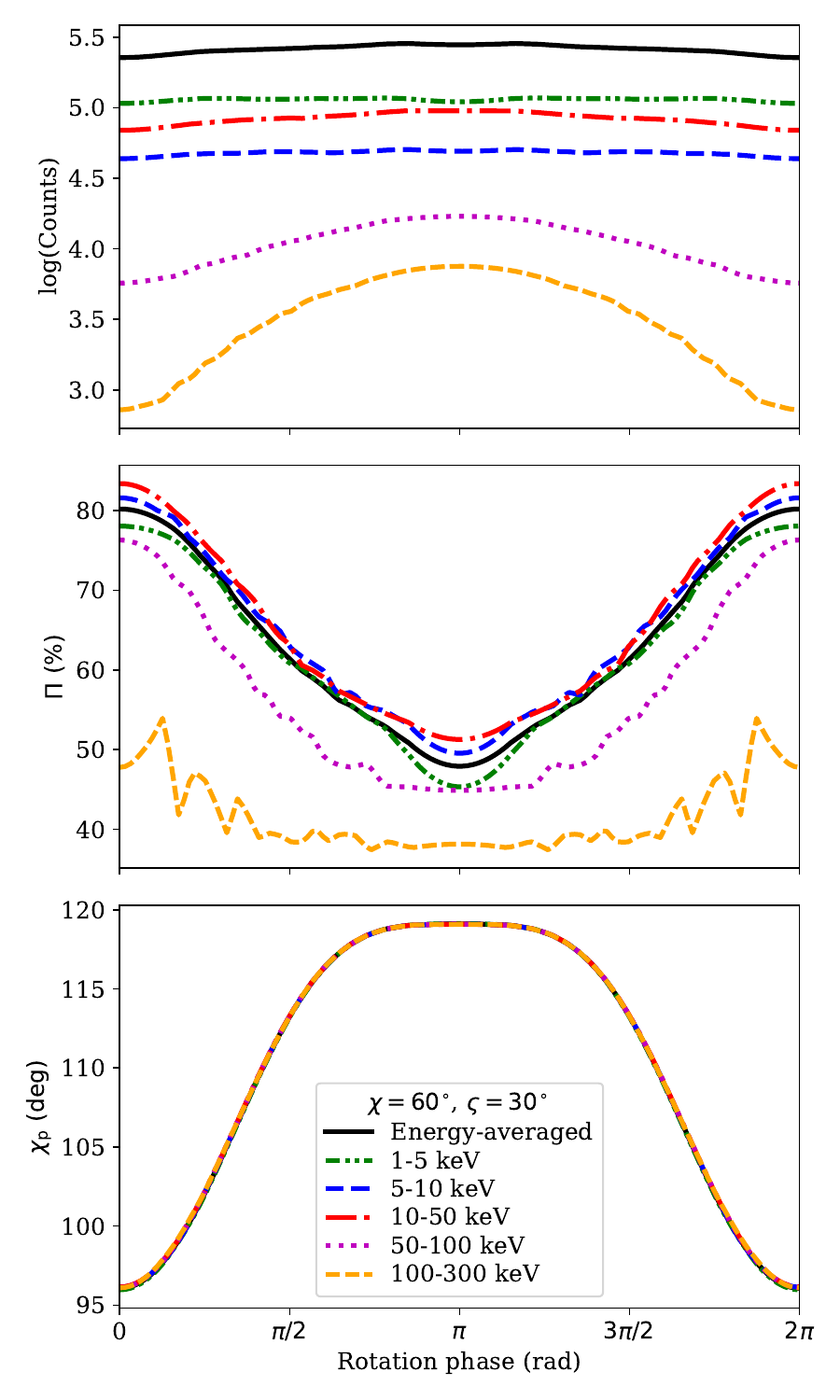}
	\includegraphics[width=5.5cm]{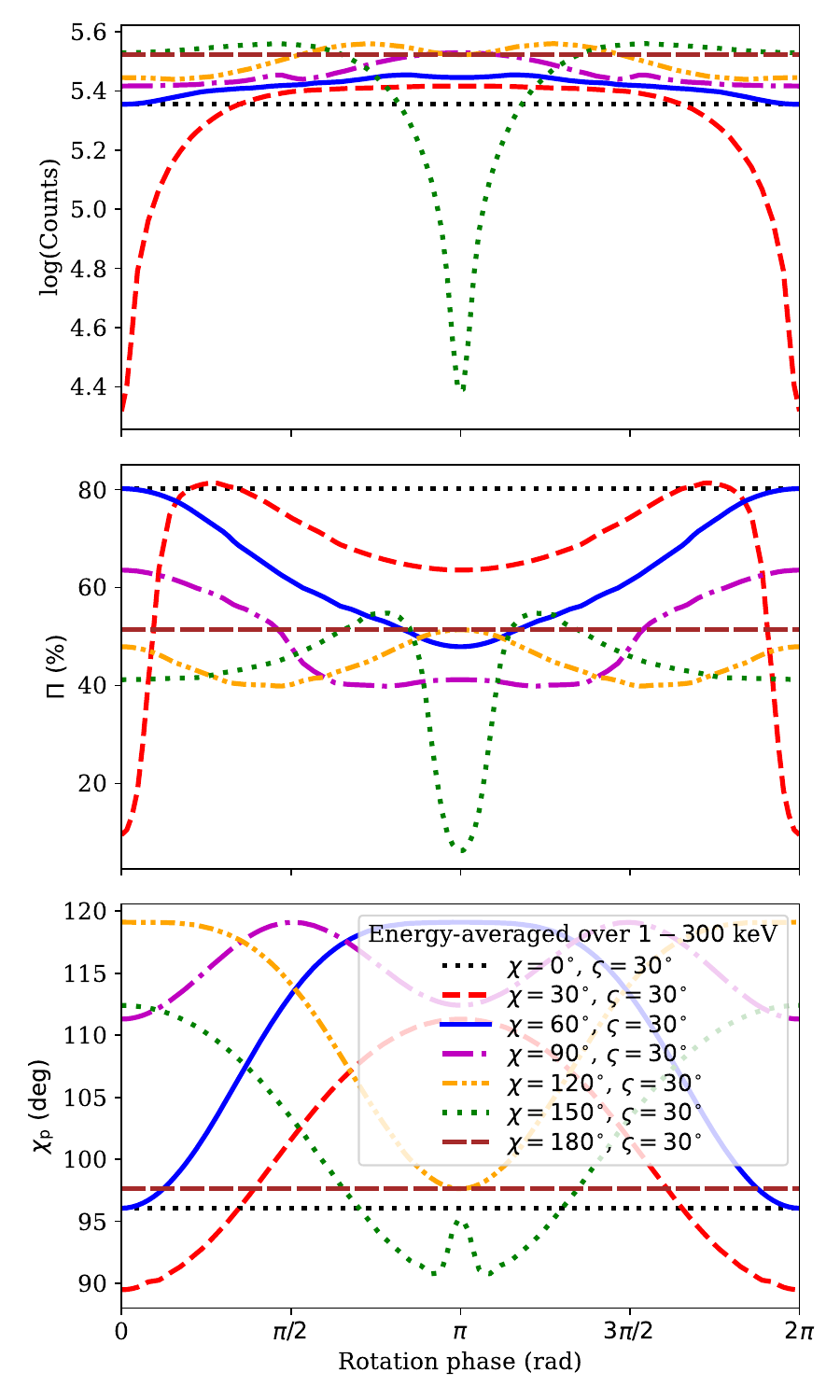}
	\includegraphics[width=5.5cm]{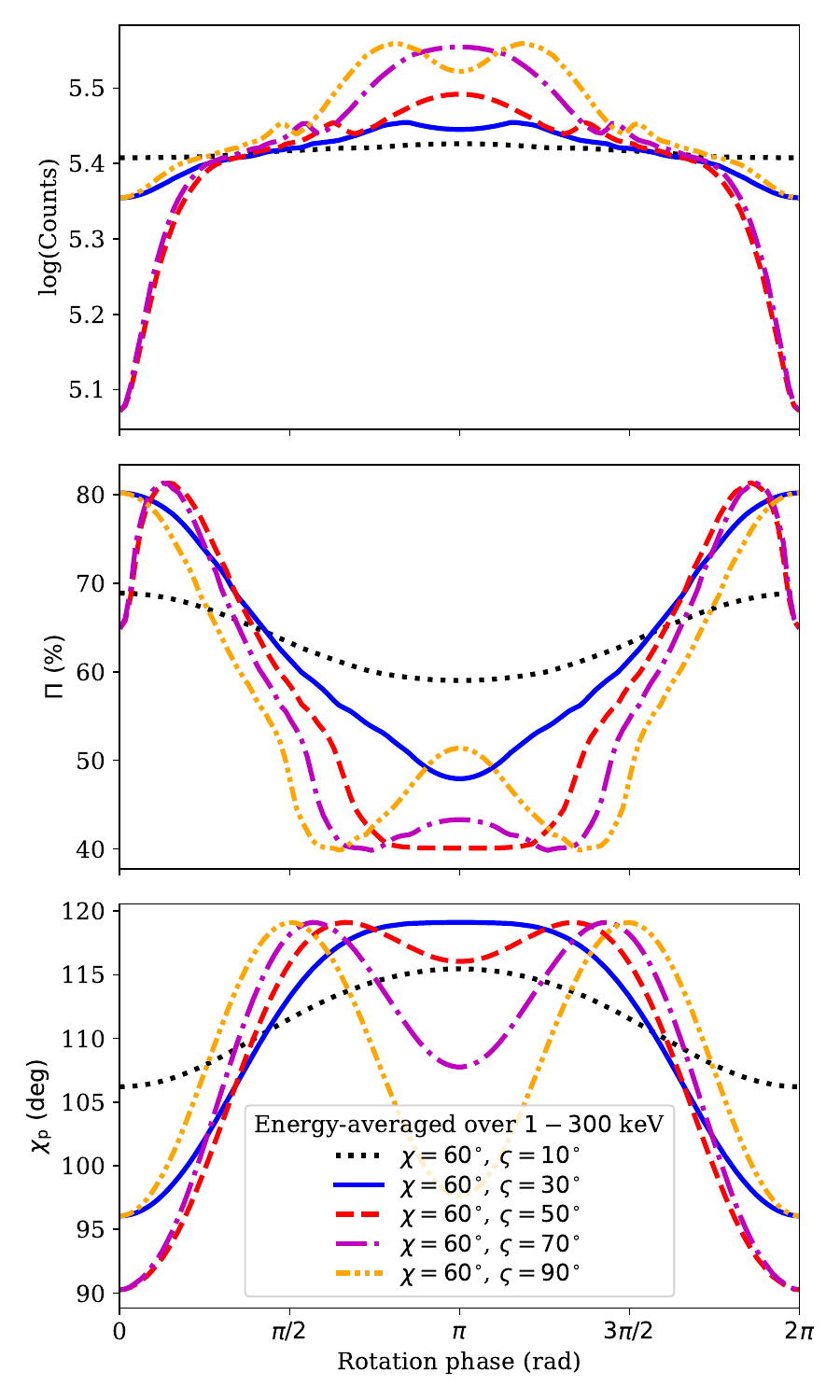}
	\caption{Pulse profiles (top panels), phase-resolved PDs ($\Pi$; middle panels), and phase-resolved PAs ($\chi_{\rm p}$; bottom panels) in different energy intervals $\varepsilon$ and angles $\chi$ as well as $\varsigma$ that the spin axis makes with the LOS as well as magnetic axis, respectively.}
	\label{fig:fig2_resolved}
\end{figure*}

\subsection{Results}
\label{subsec:results2}
To do a MC simulation, we randomly generate $10^8$ seed photons according to Section \ref{subsubsec:seed}. For each seed photon, its propagation state, i.e., polarization, direction, and energy is determined according to Section \ref{subsubsec:afterscattering}. If the photon can escape, its polarization evolution is determined according to Section \ref{subsec:polarization}. 
For each outgoing photon, its initial energy and final polarization mode $s$, direction $\theta_k$ as well as $\phi_k$, and energy $\varepsilon$ along Stokes parameters $I_i$, $U_i$, as well as $Q_i$ are stored.

\subsubsection{Phase-averaged Spectra and Polarization Properties}
\label{subsubsec:phase-averaged}
The FXRB duration $\tau_{\rm FXRB}\sim0.3$ s is much shorter than the spin period $P_{\rm s}\sim3.24$ s of SGR 1935 \citep{isr16}.
For the sake of simplicity, taking the star as an aligned rotator,
one can simulate a phase-averaged ($\phi_k$-averaged) spectrum to reproduce the time-averaged spectrum of the FXRB since the axial symmetry 
with respect to the azimuth $\phi_k$ of outgoing photons.
For an aligned rotator, $\varsigma=0$ so that $\chi=\theta_k$ and the rotation phase $\lambda=\phi_k$ (see Equation (\ref{eq:rotation})).
In this case, the photons collected within the sky patch ($\theta_k$, $\phi_k$)\footnote{When the LOS sweeps a sky patch ($\theta_k$, $\phi_k$), we set that the photons within $|\theta-\theta_k|\lesssim 1^{\circ}$ are collected.} sorted according to the energy make a phase-averaged spectrum.
We find one group of parameter values in Table \ref{tab:bestfit} can well reproduce the FXRB spectrum by MC simulation, see the left top panel of Figure \ref{fig:fig2_spec}. Within which the orange dashed line is the spectral distribution of the seed photons and the red solid line is the spectrum of the outgoing radiation after RCS.

All simulated phase-averaged spectra and polarization results are illustrated in Figure \ref{fig:fig2_spec}.
In the case of the good fitting of the FXRB spectrum with a goodness of reduced $\chi^2=23/27$, the LOS makes with the magnetic axis by a magnetic colatitude $\theta_k=40^{\circ}$.
The corresponding PD values vary with energy are displayed with the red solid line in the left middle panel, ranging from $\sim40-80\%$ in an energy band $1-300$ keV.
The right middle panel, which is the contour plot to the left middle one, shows that the PDs in low energies ($<50$ keV) seem to be averagely higher than those in high energies ($>50$ keV). This can be easily understood since the outgoing radiation with a higher energy is from the highly polarized seed photons by more multiple RCS 
and each single RCS has a specific probability of mode conversion to reduce the PD of the radiation, see Section \ref{subsubsec:afterscattering}. 
Moreover, the PDs (right middle panel) as well as the spectra (right top panel) also vary with the magnetic colatitude,
i.e., the larger magnetic colatitudes, on average, the lower PDs and the more Comptonized spectra. 
This reflects our choice for the charges applied to RCS which are the unidirectional flow electrons moving from the north to the south pole, 
as noticed in \cite{nob08} for quiescent persistent X-ray emission. 
However, it should be also noted that the largest depolarization and Comptonization does not occur when viewing the south pole but when viewing the star southern hemisphere at an intermediate angle because of the low particle density near the poles, also as noticed in \cite{nob08}.

While the PAs are nearly independent of energy, see the bottom panels of Figure \ref{fig:fig2_spec}. This is because: (a) The outgoing electric vector of each photon is influenced by only the magnetic field direction at the adiabatic radius $r_{\rm a}$ that is not strongly sensitive to energy, see Equation (\ref{eq:r_a}). (b) The outgoing radiation is still E-mode dominated in the whole energy range because an E-mode seed photon has a greater probability to retain its initial polarization state as seen from the cross-sections of RCS \citep[e.g.,][]{nob08}. On the contrary, the PAs vary with magnetic colatitude and are symmetric about the equator as clearly demonstrated in the right bottom panel. 
These results are somewhat analogous to those in \cite{fer11} and \cite{tave14} for quiescent persistent X-ray emission. 

\subsubsection{Phase-resolved Polarization Properties}
\label{subsubsec:phase-resolved}
When the star is a misaligned rotator, the coordinates ($\theta_k$, $\phi_k$) in the system ($r$, $\theta$, $\phi$) (or equivalently ($\Theta_m$, $\Phi_m$) in the system ($r$, $\Theta$, $\Phi$), see Figure \ref{fig:coordinates}), which represent the intersection of the LOS sweeping across the sky, are impacted by the rotation phase $\lambda$, see Equation (\ref{eq:rotation}).
The angles $\chi$ and $\varsigma$ that the spin axis makes with the LOS and magnetic axis cannot be well determined
from the short-duration light curve of the FXRB,
so one can arbitrarily choose the values of $\chi$ and $\varsigma$. At each group of given $\chi$ and $\varsigma$, one can obtain the corresponding pulse profile and phase-resolved polarization properties in given energy intervals, 
by a bilinear interpolation to the number of counts and polarization properties in sky patches via Equation (\ref{eq:rotation}). For further detail can refer to \cite{nob08}. 
The final results are exhibited in Figure \ref{fig:fig2_resolved}, within which the
pulse profiles (top panels) and phase-resolved PDs (middle panels) vary with energy, $\chi$, and $\varsigma$. Conversely, the PAs (left bottom panel) hardly rely on energy, 
in accordance with those in phase-averaged cases.
Furthermore, the pulse profiles, PDs, and PAs (center column panels) all vary with rotation phase, except for $\chi=0^{\circ}$ (parallel) and $180^{\circ}$ (anti-parallel) in which the visible part of the emitting region are invariable during rotation, as imagined from the right panel of Figure \ref{fig:coordinates}.
Moreover, either the pulse profiles or polarization properties (PDs and PAs) are symmetric about the rotation phase at $\pi$, which are akin to those simulations in \cite{tave14} for the quiescent persistent X-ray emission of the AXP 1RXS J170849.0$-$400910. 

\subsection{Caveats}
\label{subsec:caveats_rcs}
The first caveat is that the electron recoil during the RCS. When the photon energy $\varepsilon\ll m_e c^2/\gamma$, the cross-sections used in above RCS are safe \citep{nob08}.
From the result of Table \ref{tab:bestfit}, the Lorentz factor of electrons $\gamma\sim\mathcal{O}(1)$, so for the majority of photons with energy in the range of 10$-$100 keV, the cross-sections without the electron recoil treatment are safe. While for the minority of photons with energy 100$-$300 keV, the cross-sections are moderately safe and the corresponding results are marginally trustworthy. The fluctuations in both the PA variation for $\theta_k=10^{\circ}$ and $>100$ keV in the left bottom panel of Figure \ref{fig:fig2_spec}, and the phase-resolved PD for $100-300$ keV in the left middle panel of Figure \ref{fig:fig2_resolved}, actually reflect this conclusion.

The second one is the cutoff at energy $\sim200$ keV appearing in the case $\theta_{k}=10^{\circ}$, see the left panels of Figure \ref{fig:fig2_spec}. This cutoff is due to the seed photons above $\sim200$ keV cannot be upscattered to a higher energy via RCS because their escaping radii \citep[see the equation (26) of][]{fer11} go deep inside the emitting surface of the trapped fireball $r=R_{\max}\sin^2\theta$ where $R_{\max}=2R_{\rm NS}$ from the fitting result of Table \ref{tab:bestfit}.

The third one is that we used an axisymmetric, self-similar, globally twisted dipole magnetosphere for simplicity in this section, the twisted field may be also local and like a corona as in \cite{bel07}.

\section{Quasi-polar Trapped-Expanding Fireball vs. Non-polar Trapped Fireball}
\label{sec:quasi-polar}
The two extraordinary features of the FXRB mentioned in Section \ref{sec:extraordinary} 
can be naturally interpreted with a locale dichotomy \citep{younes20b,younes21}.
In this interpretation, the high $E_{\rm p}$ of the FXRB is owing to the less X-ray attenuation by photon splitting near quasi-polar regions \citep[e.g.,][]{hu19}.
While \cite{ioka20} and \cite{wada23} suggested that the high $E_{\rm p}$ can be naturally caused by a trapped-expanding fireball
along open magnetic field lines regardless of the photon splitting. 
This is because the observed temperature $T_{\rm obs}=\Gamma T \sim T_0$ \citep{mes00}
maintains nearly constant during the adiabatic expansion of the fireball, different from a trapped fireball relevant to an OXRB.

For the second feature over the FXRB aligning with the principal peak of the persistent emission 
that is usually treated as a hot spot on the NS surface,
there is a possibility to account for its origin as pointed out by \cite{younes20b}.
That is the hot spot originates from an internal dissipation.
This would naturally lead to the hot spot locating in a magnetic pole where the heat conduction upward from the crust is more efficient, 
because the field lines over there are oriented vertically, than a non-polar region where the field lines are oriented horizontally.
As a result, the second feature of the FXRB is also well fulfilled in a quasi-polar locale. 
This is strongly supported by the radio pulsar radiation anti-aligning with the X-ray pulsation profile for SGR 1935 recently reported by \cite{zhu23}.

Accordingly, in this section we think that the FXRB is from a trapped-expanding fireball along the open field lines at a magnetic pole, while those OXRBs are from trapped fireballs enclosed in the closed field lines at non-polar regions, regardless of the photon splitting.

\subsection{Dynamics and Radiation from an Expanding Fireball}
\label{subsec:expanding}
For simplicity, we assume that an initial trapped fireball only consists of $e^{\pm}$ pair-photon plasma containing no baryons, expanding along the open field lines at a magnetic pole.
The basic picture is elaborated as in \cite{ioka20} and \cite{wada23}: at the base of the open field lines, 
a small-scale trapped fireball could be formed due to a crustal deformation or fracture, and diffusively supplies $e^{\pm}$ pair-photon to the flux tubes of the open field lines.
As the fireball expands, its Lorentz factor $\Gamma$, comoving temperature $T$,  lateral size $\ell$, and pair number density $n_{\pm}$ evolve as\footnote{In this section, we use as many of the notations from \cite{wada23} as possible.}:
\begin{equation}
	\Gamma = \Gamma_0\tilde{r}^{3/2} = \tilde{r}^{3/2},
	\label{eq:Gam} 
\end{equation}
\begin{equation}
	T = T_0\tilde{r}^{-3/2},
	\label{eq:T} 
\end{equation}
\begin{equation}
	\ell = \ell_0\tilde{r}^{3/2},
	\label{eq:l} 
\end{equation}
and
\begin{equation}
	n_{ \pm}(T, B) =\begin{cases} 4\left(\frac{m_e^2 c^2}{2 \pi \hbar^2}\right)^{3 / 2} B T^{1 / 2} \exp \left(-1/T\right), \\
		{\rm if}~m_e  c^2~\&~\hbar \omega_e (1) \gg m_e c^2 T   \\
		4\left(\frac{m_e^2 c^2}{2 \pi \hbar^2}\right)^{3 / 2} T^{3 / 2} \exp \left(-1/T\right), \\
		{\rm if}~m_e c^2 \gg m_e c^2 T \gg \hbar \omega_e (1), \end{cases} 
	\label{eq:n_pm}
\end{equation}
where $\tilde{r}=r/R_{\rm NS}$ in which $r(>R_{\rm NS})$ is the fireball radius above the stellar surface, $\Gamma_0=1$, $T_0$, and $\ell_0$ are initial conditions, 
and $\hbar \omega_e (1)=m_e c^2(\sqrt{1+2B}-1)$ is the energy of the first Landau level \citep{thom95}.  
Furthermore, the comoving temperature and magnetic field have been normalized by $T \rightarrow \frac{k_{\rm B} T}{m_e c^2}$ and $B \rightarrow \frac{B}{B_Q}$ where $B_Q=\frac{m_e^2 c^3}{e\hbar}$ is the critical field strength and
\begin{equation}
	B \approx B_0\tilde{r}^{-3}
	\label{eq:B}
\end{equation}
at the polar region for a pure dipole field (see Equation (\ref{eq:B_rtp}) in case of $\theta\rightarrow0$)\footnote{While in the polarization study, the $\theta$ dependence in magnetic field $\boldsymbol{B}$ is important, see Equation (\ref{eq:QUv}).}, 
where $B_0$ is the normalized surface field strength at the poles. We use these normalized $T(T_0)$ and $B(B_0)$ throughout this section.

Photons finally escape from the fireball in two ways.
One way is photons escape longitudinally, when the optical depth in the direction parallel to the magnetic field lines becomes $\tau_{\|}<1$. 
The other is photons diffuse out laterally via diffusion, when the diffusion timescale of photons perpendicular to the magnetic field lines 
and the dynamical timescale satisfy $t_{\rm diff}<t_{\rm dyn}$. The optical depth $\tau_{\|}$ and $t_{\rm diff}/t_{\rm dyn}$ are given by
\begin{equation}
	\begin{aligned}
		& \tau_{\|}=\tau_{ \pm 0}\tilde{r}^{\delta^{\prime}} \exp \left(-A/T_0\right) \\
		& \frac{t_{\mathrm{diff}}}{t_{\mathrm{dyn}}}=\tau_{ \pm 0} \theta_0^2 \tilde{r}^{\zeta^{\prime}} \exp \left(-A/T_0\right),
	\end{aligned}
	\label{eq:tau}
\end{equation} 
where $A=\tilde{r}^{3/2}-1$, $\theta_0=\ell_0/R_{\rm NS}$, $t_{\rm dyn}=r/(c\Gamma)$ is the dynamical timescale, and $t_{\rm diff}=n_{ \pm}(T, B)\sigma(T, B) \ell^2/c$ is the diffusion timescale in which the scattering cross-section is described as
\begin{equation}
	\sigma(T, B)=\begin{cases}\frac{4 \pi^2}{5} T^2 B^{-2} \sigma_{\rm T}, & {\rm E{\text -}mode} \\
		\sigma_{\rm T}, & {\rm O{\text -}mode},\end{cases}
	\label{eq:sigma} 
\end{equation}
where $\sigma_{\rm T}$ is the Thomson scattering cross-section.
$\delta^{\prime}$, $\zeta^{\prime}$, and $\tau_{ \pm 0}$ are separately shown in Equations (\ref{eq:delta_p}), (\ref{eq:zeta_p}), and (\ref{eq:tau_pm0}) for three cases\footnote{One can refer to the first, second, and fifth rows in the table 3 of \cite{wada23}. The RD/O-mode/lL case means that the fireball is in radiation-dominated phase, the suppression for E-mode photons does not occur, and the $e^{\pm}$ pairs only occupy the lowest Landau level. The RD/O-mode/hL is the same as the first case except that the $e^{\pm}$ pairs occupy the higher Landau levels. While the RD/E-mode/lL is the same as the first case except that the suppression for E-mode photons occur.}: RD/O-mode/lL, RD/O-mode/hL, and RD/E-mode/lL.

One can obtain the longitudinal escaping radius $r_{\|,\pm}$
and lateral-diffusion radius $r_{\rm diff,\pm}$ by solving $\tau_{\|}=1$ and $t_{\rm diff}/t_{\rm dyn}=1$ according to Equation (\ref{eq:tau}), respectively.
From the parameter values $B_0B_Q=B_{\rm p}\sim2\times10^{14}$ G for SGR 1935, and tentative initial temperature $m_ec^2T_0\sim80$ keV and $\theta_0=0.01$ (i.e., $\ell_0=10^4$ cm) for the fireball as adopted in \cite{ioka20}, for all three aforementioned cases one would get that: (a) the values of $r_{\|,\pm}$ and $r_{\rm diff,\pm}$\footnote{Where $r_{\|,\pm}>r_{\rm diff,\pm}$ is always satisfied for these three cases: RD/O-mode/lL, RD/O-mode/hL, and RD/E-mode/lL.} are both smaller than the scattering-suppression radius 
$r_E=\left(4 \pi^2 / 5\right)^{-1 / 3} T_0^{-2 / 3} B_0^{2 / 3} R_{\rm NS}\sim4.7R_{\rm NS}$ \citep[cf. the table 1 of][]{wada23}, 
(b) $\hbar \omega_e(1)\gg m_e c^2T$ at $r_{\rm diff,\pm}$. 
The former result signifies that the firstly escaping photons should be E-mode dominated and the relative optical depth $\tau_{\|}$ as well as $t_{\rm diff}/t_{\rm dyn}$
only based on the RD/E-mode/lL case are applicable. 
If so, one would have $r_{\|,\pm}=2.4R_{\rm NS}$ and $r_{\rm diff}=1.9R_{\rm NS}$ by solving $\tau_{\|}=1$ and $t_{\rm diff}/t_{\rm dyn}=1$.
The latter result indicates that the $e^{\pm}$ pairs still only occupy the first Landau level when the E-mode photons firstly escape and the O-mode photons subsequently escape.
In this case, based on the $\delta^{\prime}$, $\zeta^{\prime}$, and $\tau_{ \pm 0}$ values of for the RD/O-mode/lL case, the longitudinal escaping radius $r_{\|,\pm}=2.5R_{\rm NS}$ and lateral-diffusion radius $r_{\rm diff,\pm}=2.1R_{\rm NS}$ for the subsequently escaping O-mode photons can be obtained by solving $\tau_{\|}=1$ and $t_{\rm diff}/t_{\rm dyn}=1$.

Since $r_{\|,\pm}>r_{\rm diff,\pm}$\footnote{Here $r_{\|,\pm}$ is actually the photospheric radius without lateral expansion, while $r_{\rm diff,\pm}$ is the photospheric radius with lateral expansion \citep[cf. the section 3.1 in][]{wada23}.}, before photons escape in the direction longitudinal to the magnetic field lines, the lateral size of the fireball increases from $\ell(r_{\rm diff,\pm})$ to $\ell_{\rm sphe}$, and the temperature rapidly decreases from $T(r_{\rm diff,\pm})$ to $T_{\rm sphe}$ at which photons begin to escape longitudinally.
$T_{\rm sphe}$ is therefore the comoving photospheric temperature, which can be determined by \citep{wada23}
\begin{equation}
	n_{ \pm}(T_{\rm sphe}, B(r_{\rm diff,\pm}))\sigma(T_{\rm sphe}, B(r_{\rm diff,\pm}))\frac{r_{\rm diff,\pm}}{\Gamma(r_{\rm diff,\pm})}=1.
	\label{eq:T_sphe} 
\end{equation}
The lateral size of the photospheres can be then evaluated as
\begin{equation}
	\ell_{\rm sphe}=\ell(r_{\rm diff,\pm}) \left( \frac{T(r_{\rm diff,\pm})} {T_{\rm sphe}} \right)^2.
	\label{eq:l_sphe} 
\end{equation}
From known $r_{\rm diff,\pm}$ and Equations (\ref{eq:Gam}), (\ref{eq:n_pm}), (\ref{eq:B}), and (\ref{eq:sigma}), one solves Equation (\ref{eq:T_sphe}) to
get the comoving E-mode photospheric temperature $m_e c^2 T_{\rm sphe}\sim22$ keV and the O-mode $\sim22$ keV. 
At these temperatures, their relative Lorentz factors are $\Gamma(r_{\rm diff,\pm})\sim2.7$ for the E-mode photosphere and $\sim3.1$ for the O-mode, 
calculated from Equation (\ref{eq:Gam}). Their lab-frame photospheric temperatures can thus be derived by
\begin{equation}
	T_{\rm lab}= \Gamma\left(r_{\rm diff, \pm}\right)T_{\rm sphe}.
	\label{eq:T_lab}
\end{equation}

It should be taken note of the half-opening angle of the fireball $\theta_{\rm j}=\frac{1}{2}\frac{\ell_{\rm sphe}}{r_{\rm diff, \pm}}\sim0.01$ rad 
from the radial and lateral expansions of either the E-mode or O-mode photons.
Because of a relativistic beaming, the E-mode photons emitted within a cone with a beaming angle of $\theta_{\rm b} \sim 1/\Gamma(r_{\rm diff, \pm})\sim0.37$ rad and the O-mode photons emitted within $\sim0.32$ rad. Both of them are much larger than that of the fireball. 
Accordingly, a distant observer can view the radiation from the whole fireball photosphere during it sweeps across the LOS
\footnote{This is different from the highly beamed ``lighthouse'' radio pulsar radiation, because only the radiation within a narrow cone that is beamed on the LOS, 
	rather than the radiation from the whole polar cap, can be viewed by a distant observer at a moment in rotation phase. For this narrow radiation, if it consists of single mode 
	(either E-mode or O-mode) photons, it will be highly polarized when it sweeps across the LOS 
	and its PA variation with phase is generally described by the rotating vector model \citep{rad69,lyne88}. 
	While looking at FRBs, they seem to be also originated from a polar cap \citep{kumar20} but occupy fragmented quasi-tangential regions \citep{liu24}. 
	If this is the case, they should look like the lotus seeds in the lotus seed head if the polar cap looks like a lotus seed head.}. 
If the fireball photosphere is presumed to be steady during its short duration, 
the radiation from such a photosphere confined within a beaming angle $\theta_{\rm b}$ is somewhat similar to that radiation coming from a polar hot spot on an NS surface. 

\begin{deluxetable*}{cccccccccc}
	\label{tab:efball}
	\tablecaption{Best-fit Parameter Values and Derived E-mode and O-mode Photospheric Properties}
	\tablehead{\multicolumn{3}{c}{Best-fit Parameter Values} & & \multicolumn{6}{c}{Derived Values}
	}
    \startdata
		\cline{1-3} \cline{5-10} \\
		$\ell_0$ & $T_0$ & Goodness & Photosphere & $r_{\rm diff,\pm}$ & $\Gamma(r_{\rm diff,\pm})$ & $T_{\rm sphe}$ & $T_{\rm lab}$ & $\theta_{\rm j}$ & $\theta_{\rm b}$ \\
		(cm) & (keV) & ($\chi^2/{\rm dof}$) & ($R_{\rm NS}$) & & (keV) &	(keV) & (rad) & (rad) \\
		\hline
		\multirow{2}{*}{$(4.0^{+0.1}_{-0.1})\times10^4$} & \multirow{2}{*}{$41^{+1}_{-1}$} & \multirow{2}{*}{48/31} & E-mode & 1.1 & 1.12 & 23.5 & 26.3 & 0.013 & 0.90 \\
		& & & O-mode & 1.4  & 1.58 & 23.0 & 36.3 & 0.007 & 0.63
	\enddata
	\tablenotetext{Notes.}
	{\\ $\ell_0$ and $T_0$ are the inital lateral size and comoving temperature of the fireball, respectively. $r_{\rm diff,\pm}$ is the lateral-diffusion (photospheric) radius, 
		$\Gamma(r_{\rm diff,\pm})$ is the photospheric Lorentz factor at radius $r_{\rm diff,\pm}$, $T_{\rm sphe}$ is the comoving photospheric temperature and its lab-frame one is
		$T_{\rm lab}=\Gamma(r_{\rm diff,\pm}) T_{\rm sphe}$, 
		$\theta_{\rm j}=\frac{1}{2}\frac{\ell_{\rm sphe}}{r_{\rm diff,\pm}}$ is the half-opening angle of the fireball, 
		and $\theta_{\rm b}\sim1/\Gamma(r_{\rm diff,\pm})$ is the beaming angle of the radiation. }
\end{deluxetable*}

\subsection{Spectrum and Polarization of Radiation}
\label{subsec:fb_spec}
Learning from the spectrum and polarization research for X-ray radiation of a polar hot spot \citep{per08,van09}, 
one may describe the spectrum of the total radiation attributed to both the E-mode and O-mode photospheres of a polar expanding fireball by a double blackbody function with different temperatures\footnote{The time interval between the E-mode and O-mode photospheric radiation is extremely short ($\ll1$ ms) due to a short timescale with order of $\sim t_{\rm dyn}+t_{\rm diff}<2t_{\rm dyn}=\frac{2r}{c\Gamma}$ because of $r$ being of order $R_{\rm NS}$, so the total radiation at one moment are attributed to both the E-mode and O-mode photospheres.}.
Through the E-mode and O-mode lab-frame photospheric temperatures from Equation (\ref{eq:T_lab}), one can construct a photon spectrum as
\begin{equation}
	\begin{aligned}
		n=&n_{\rm E}+n_{\rm O} \\
		=&\frac{2\pi\varepsilon^2 }{c^2h}\frac{1}{\exp[{\varepsilon/(m_e c^2T_{\rm lab,E})}]-1} \frac{r_{\rm diff,\pm,E}^2}{D_{\rm L}^2} \\
		&+\frac{2\pi\varepsilon^2 }{c^2h}\frac{1}{\exp[{\varepsilon/(m_e c^2T_{\rm lab,O})}]-1} \frac{r_{\rm diff,\pm,O}^2}{D_{\rm L}^2}.
	\end{aligned}
	\label{eq:n} 
\end{equation} 

To obtain the observed time-averaged spectrum, one should consider the polar expanding fireball geometry that is somewhat like a polar hot spot.
In the spherical coordinate system ($r$, $\Theta$, $\Phi$) with the $Z$-axis along the LOS, 
the geometry is illustrated in the right panel of Figure \ref{fig:coordinates} and the visible fireball (i.e., its photosphere) is expressed by the following conditions \citep{per08}:
\begin{equation}
	\begin{gathered}
		0\leqslant\Theta \leqslant \theta_{\rm j}, 0\leqslant \Phi\leqslant 2\pi, \quad \text { if } \quad \Theta_m=0, \\
		\Theta \leqslant \Theta_*\left(\Theta_m, \theta_{\rm j}, \Phi\right), 0\leqslant \Phi\leqslant 2\pi, \quad \text { if } \quad \Theta_m \neq 0, \Theta_m \leqslant \theta_{\rm j}, \\
		\left\{\begin{array}{l}
			\Theta_m-\theta_{\rm j} \leqslant \Theta \leqslant \Theta_m+\theta_{\rm j}, \\
			\Phi_{\rm h}  \leqslant \Phi \leqslant 2 \pi-\Phi_{\rm h},
		\end{array} \text { if } \quad \Theta_m \neq 0, \theta_{\rm j} < \Theta_m \leqslant \theta_{\rm b},\right.  \\
		\text{Receiving zero radiation since no radiation outside $\theta_{\rm b}$}, \\
		\text{if}  \quad \Theta_m > \theta_{\rm b},
	\end{gathered}
	\label{eq:Theta-Phi}
\end{equation}  
where $\Theta_*\left(\Theta_m, \theta_{\rm j}, \Phi\right)$ is computed by numerical solution of
\begin{equation}
	\cos \theta_{\rm j}=\sin \Theta_* \sin \Theta_m \cos \Phi+\cos \Theta_* \cos \Theta_m,
\end{equation}
and \begin{equation}
	\Phi_{\rm h}=\arccos \left(\frac{\cos \theta_{\rm j}-\cos \Theta_m \cos \Theta}{\sin \Theta_m \sin \Theta}\right).
\end{equation}

\begin{figure}
	\centering
	\includegraphics[width=\columnwidth]{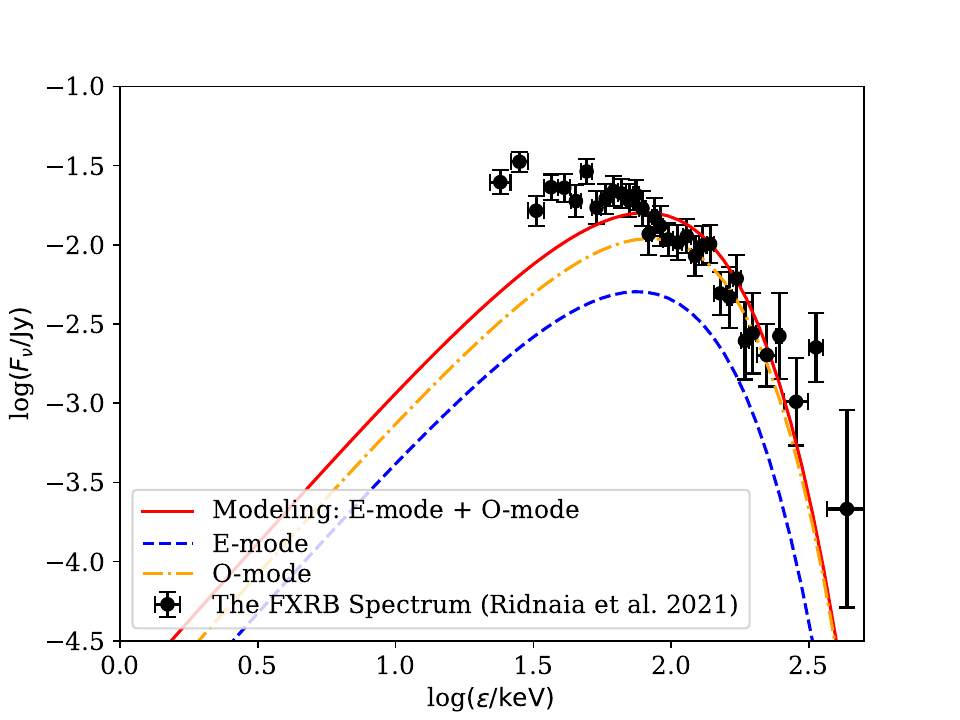}
	\caption{The best fit for the spectrum of the FXRB associated with FRB 20200428D within the trapped-expanding fireball model. The dash blue and dotted orange lines represent E-mode and O-mode photospheric radiation, respectively. While the solid red line represents the total radiation.}
	\label{fig:efball}
\end{figure}

\begin{figure*}
	\centering
	\includegraphics[width=8.5cm]{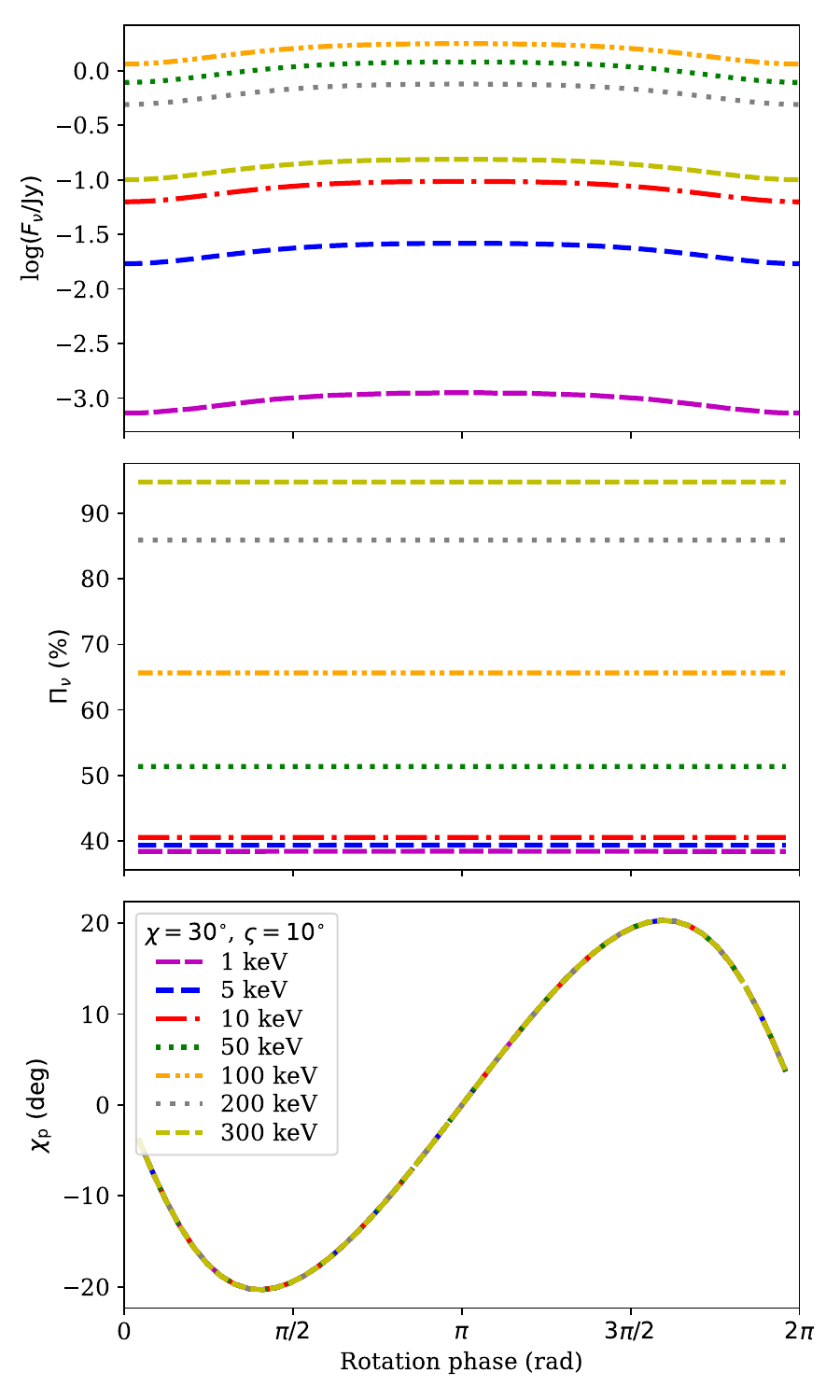}
	\includegraphics[width=8.5cm]{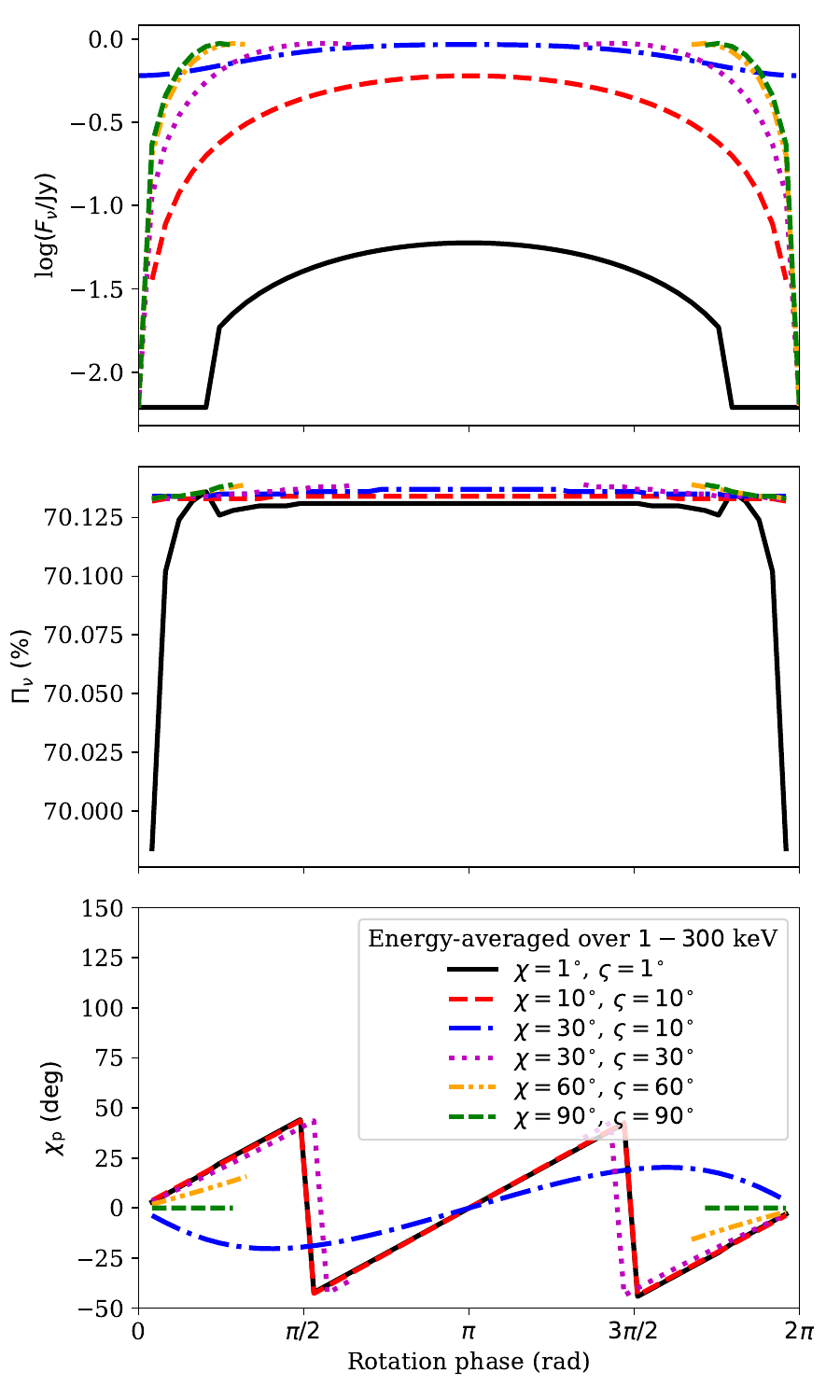}
	\caption{Pulse profiles ($F_{\nu}$; top panels), phase-resolved PDs ($\Pi_{\nu}$; middle panels), and phase-resolved PAs ($\chi_{\rm p}$; bottom panels) in different energies $\varepsilon$ and angles $\chi$ as well as $\varsigma$ that the spin axis makes with the LOS as well as magnetic axis, respectively.}
	\label{fig:ef_Pol}
\end{figure*}

According to the method in \cite{tave15},
the observed total photon flux is obtained by
\begin{equation}
	F_{\nu}=I_{\nu}= \int d \Phi \int d u^2 \left(n_{\rm E}+n_{\rm O}\right),
	\label{eq:Fv}
\end{equation}
where $u=\sin\Theta$, the integral ranges of $\Theta$ and $\Phi$ are in Equation (\ref{eq:Theta-Phi}), for simplicity we have neglected the strong gravity on the relativistic ray bending, ray redshift, and stellar magnetic field throughout this section.
In a similar way, the total Stokes parameters $Q_{\nu}$ and $U_{\nu}$ are given by
\begin{equation}
	\begin{aligned}
		& Q_{\nu}= \int d \Phi \int d u^2\left(n_{\rm E}-n_{\rm O}\right) \cos (2 \alpha) \\
		& U_{\nu}= \int d \Phi \int d u^2\left(n_{\rm E}-n_{\rm O}\right) \sin (2 \alpha),
		\label{eq:QUv}
	\end{aligned}
\end{equation}
where $\cos (2 \alpha)$ and $\sin (2 \alpha)$ can be calculated from $\cos\alpha=-B_Y/\sqrt{B_X^2+B_Y^2}$ in which $B_X$ and $B_Y$ are in Equation (\ref{eq:B_XYZ}).
Since the magnetic field direction at the adiabatic radius not at that field direction within the adiabatic region eventually influences the outgoing electric vector of each photon,
$B_X$ and $B_Y$ are fixed at the adiabatic radius $r_{\rm a}\simeq4.8(B_{\rm p}/10^{11}{\rm G})^{2/5}(\varepsilon/1{\rm keV})^{1/5}R_{\rm NS}$ 
which is suitable for a pure dipole field \citep{tave15}.
Obviously, the Stokes parameters are intrinsically dependent on $\chi$, $\varsigma$, and $\lambda$. So the observed time-averaged spectrum (i.e., the phase-averaged spectrum) of the FXRB can be estimated by $F_{\nu,\rm ave}= \frac{1}{2\pi \tau_{\rm r}}\int_0^{2\pi \tau_{\rm r}} F_{\nu}(\lambda) d\lambda$ where $\tau_{\rm r}=\tau_{\rm FXRB}/P_{\rm s}$.

\subsection{Results}
\label{subsec:efball-results}
To fit the FXRB spectrum by Markov chain Monte Carlo (MCMC) algorithm, we firstly take the initial lateral size $\ell_0$ and comoving temperature $T_0$ of the fireball as two free parameters, subsequently get the lateral-diffusion radius $r_{\rm diff,\pm}$ and the comoving (lab-frame) temperatures $T_{\rm sphe}$($T_{\rm lab}$) of E-mode as well as O-mode photospheres based on the dynamical formulae in Section \ref{subsec:expanding}, and finally use Equations (\ref{eq:n}) and (\ref{eq:Fv}) to reach the observed spectrum. For the simplest case $\chi=\varsigma=0$ and thus $\Theta_m=0$, $F_{\nu}$ is actually the observed time-averaged spectrum because it is no longer phase-dependent. The MCMC running results are listed in Table \ref{tab:efball} and the best fitting to the FXRB spectrum is exhibited in Figure \ref{fig:efball}. From the results one can obtain that: (a) The total radiation is O-mode dominated in the whole energy band of $1-300$ keV, from the photon number flux (not the flux density $F_{\nu}$ in Figure \ref{fig:efball}) comparison between $n_{\rm O}$ and $n_{\rm E}$ (Equations (\ref{eq:n})).
(b) The fireball properties in Table \ref{tab:efball} obtained from the spectral fitting for the simplest case should be adequate to preliminarily grasp the relevant polarization properties, 
which are computed by the same formulae as Equation (\ref{eq:PD-PA}). 

When the fireball is in the case of $\chi=\varsigma=0$ and thus $\Theta_m=0$ (i.e., the LOS as well as the spin axis both are parallel to the magnetic axis), 
its relative $Q_{\nu}$ and $U_{\nu}$ are zeros and thus there is no polarization.
This is because the total radiation from the visible part of the fireball is always axisymmetric, no matter which mode radiation is dominated.
It is in agreement with the polarization result for the radiation from a trapped fireball noticed by \cite{yang15}, see the left panel of their figure 10. 

When the fireball is not in the case of $\chi=\varsigma=0$, 
the pulse profiles and polarization properties are shown in Figure \ref{fig:ef_Pol} for different energies $\varepsilon$ and various $\chi$ as well as $\varsigma$. 
One can find: 
(a) The pulse profiles ($F_{\nu}$; top panels) vary with energy, $\chi$, $\varsigma$, and rotation phase, and are all symmetric about the rotation phase in $\pi$. 
(b) The energy-sliced PD ($\Pi_{\nu}$) values vary from $\sim40-95\%$ with energy in the range of $1-300$ keV for a fixed $\chi=30^{\circ}$ and $\varsigma=10^{\circ}$ (left middle panel). 
Furthermore, the energy-averaged PDs over $1-300$ keV are nearly constant with values $\sim70\%$ for various $\chi$ and $\varsigma$ (right middle panel). 
Moreover, all PDs are nearly invariant with rotation phase.
(c) The energy-sliced PAs ($\chi_{\rm p}$) are independent of energy since the total radiation is O-mode dominated in the whole energy range of $1-300$ keV (left bottom panel). 
Additionally, the energy-averaged PAs over $1-300$ keV vary with $\chi$ and $\varsigma$ (right bottom panel) and are all axisymmetric about the rotation phase in $\pi$.
(d) The cuts in pulse profiles, PDs, and PAs (right panels) for a few cases such as $\chi=\varsigma=90^{\circ}$ ascribe to the condition of $\Theta_m > \theta_{\rm b}$ in which the observer cannot receive radiation anymore since there is no radiation outside $\theta_{\rm b}$, see Equation (\ref{eq:Theta-Phi}).

\subsection{Caveat}
\label{subsec:caveat_efball}
In this section, we address a caveat related to the observed total radiation, which results from the combined emission of the E-mode and O-mode photospheres. Within this context, we treat the E-mode and O-mode photospheres are independent. In reality, when E-mode photos begin to laterally diffuse, O-mode photons also escape due to a mode exchange from O-mode to E-mode. This is because when E-mode photons escape from the fireball, O-mode ones still interact with the pairs in the fireball, potentially changing their polarization mode upon scatterings \citep[see, e.g.,][]{her79,ven79,mes92}. However, it's worth noting that in the pair-diffusion case, sufficient conversion from O-mode to E-mode photons may not occur because the diffusion radius is very close to the photospheric radius, as mentioned in \cite{wada23}. Furthermore, the diffusion and photospheric radii of E-mode photons are also remarkably close to those of O-mode photons. These are supported by the analysis in Section \ref{subsec:expanding} and the results presented in Table \ref{tab:efball}, suggesting that once E-mode photons start escaping, O-mode photons follow suit without undergoing sufficient O-mode to E-mode conversion. Consequently, treating the E-mode and O-mode photospheres as two independent sources appears reasonable. 

\section{Synchrotron Radiation in a Relativistic Shock}
\label{sec:synchrotron}
In this model, an FXRB is thought to stem from
the incoherent synchrotron radiation from relativistically
hot electrons heated by the same shock that generates FRB \citep{lyu14,met19,mar20}.
To explore its polarization signature,
one needs to combine the dynamics of relativistic shock with the polarization nature of synchrotron radiation,
as usually done in gamma-ray bursts \citep[e.g.,][]{ghi99,gru99,sari99,gran03a,gran03b,lyut03,fan05,toma09,lan16}.

\subsection{Dynamics of Relativistic Shock}
\label{subsec:dynamics}
In a flaring magnetar scenario, magnetar suddenly injects a flare with an
isotropic energy $E$ over a short timescale $\delta t \sim10^{-4}-10^{-3}$ s
(i.e., typical duration of central engine activity or FRB duration in the lab frame),
producing a radially expanding ejecta with an initial bulk Lorentz factor $\Gamma_{\rm ej}\gg 1$.
This ultrarelativistic ejecta collides with the upstream subrelativistic ion-loaded shell ejected from a recent earlier flare.
The description of the upstream ion shell and the dynamics of the ejecta deceleration can be referred to \cite{met19} in detail,
here we briefly list the key expressions.

The upstream medium perhaps more realistically is a discrete shell whose density is expressed by
\begin{equation}
	n_{\rm ext}=\frac{3\dot{M}}{4 \pi  m_{\rm p} v_{\rm w}^3 \Delta t^2}, 
	\label{eq:n_ext}
\end{equation}
where $m_{\rm p}$ is the proton mass, $\Delta t$ is the timescale of the shell injection
(i.e., the average time interval between two bursts),
$\dot{M}$ and $v_{\rm w}$ are the injection rate and velocity of the shell, respectively.
Above equation is derived from the equations (4) and (5) in \cite{met19}.

The deceleration of the ejecta by the upstream ion shell can be described by an early phase that a reverse shock passes through
the ejecta at radius $r\ll r_{\rm dec}$ \citep[$r_{\rm dec}$ is the deceleration radius\footnote{The distance from the center of explosion where roughly half of the ejecta energy $E/2$ is transferred to the upstream medium.} in the source frame;][]{sari95}
and a late phase that a forward shock enters a self-similar evolution at $r\gg r_{\rm dec}$ \citep{blan76}.
If the upstream shell can be seen as being approximately
stationary ($\beta_{\rm w}=v_{\rm w}/c\ll 1$) in the source frame, 
the Lorentz factor of the shocked fluid with respect to the upstream shell evolves as a function of radius $r$ by
\begin{equation}
	\Gamma=\begin{cases}\Gamma_{\rm e0}r^{-1/2}=\left(\frac{E \Delta t^2 \beta_{\rm w}^3}{12\dot{M}\delta t}\right)^{1/4} r^{-1/2}, &r\ll r_{\rm dec} \\
		\Gamma_{\rm l0}r^{-3/2}=\left(\frac{17E \Delta t^2 \beta_{\rm w}^3 c}{12\dot{M}}\right)^{1/2} r^{-3/2}, &r\gg r_{\rm dec}, \end{cases}
	\label{eq:Gamma}
\end{equation}
when the equations (6), (7), (9), and (10) in \cite{met19} and above Equation (\ref{eq:n_ext}) are used.
The deceleration radius $r_{\rm dec}$ can be calculated by equating the first and second lines in Equation (\ref{eq:Gamma}),
\begin{equation}
	r_{\rm dec}=\frac{\Gamma_{\rm l0}}{\Gamma_{\rm e0}}.
	\label{eq:r_dec}
\end{equation}

The radius of shocked fluid $r$ is related to the observer frame time $t_{\rm obs}$ via
\begin{equation}
	r = \frac{\beta c}{1-\beta\cos\Theta}\frac{t_{\rm obs}}{1+z},
	\label{eq:r}
\end{equation}
where $\beta\equiv(1-1/\Gamma^2)^{1/2}$, $z$ is the redshift here,
and $\Theta$ is the inclination angle measured
from the LOS, see Figure \ref{fig:syn_geo} that is replotted from the figure 1 of \cite{gill20}. This equation reflects the equal arrival time surface.

\subsection{Spectrum and Polarization from a Globally Ordered Magnetic Field}
\label{subsec:psr}

\begin{figure}[ht]
	\includegraphics[width=8cm]{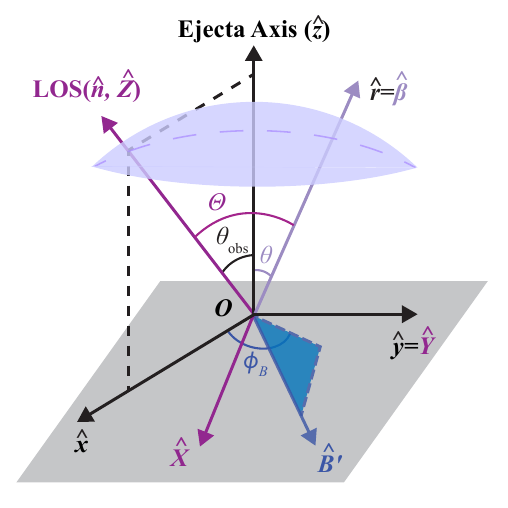}
	\includegraphics[width=8cm]{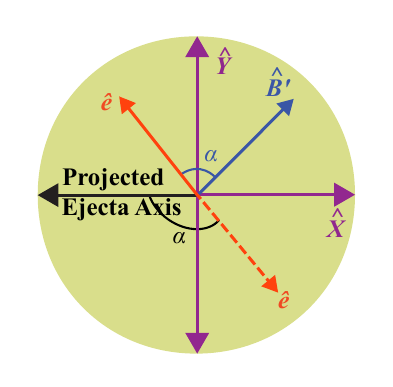}
	\caption{{\it Upper}: Illustration of the coordinate systems in which the polarization vector
		relevant to synchrotron emission is calculated. Here the local bulk velocity direction
		is $\boldsymbol{\hat\beta}=\boldsymbol{\hat r}$ and the uniform magnetic field direction is transverse to that
		with azimuthal angle $\phi_B$. The inclination angle $\Theta$ in the lab-frame is between the directions of the
		local bulk velocity and observed photon (the LOS $\boldsymbol{\hat n}$), with $\cos\Theta=\boldsymbol{\hat n}\cdot\boldsymbol{\hat\beta}$.
		{\it Lower}: The observer sees the projection of the ordered magnetic field (blue arrow)
		and polarization vector (red arrow) on the plane of the sky (shaded yellow region; orthogonal to
		the observed photon direction with wave vector $\boldsymbol{\hat k}=\boldsymbol{\hat n}$, which points out of the page).
		For an ordered magnetic field the PA $\alpha$ is measured from the
		the ordered field direction (solid arrow), otherwise $\alpha$ is measured from the projection
		of the ejecta axis (dashed arrow).}
	\label{fig:syn_geo}
\end{figure}

Owing to the resulting linear
polarization of FRB emission along the direction of the star spin axis $\boldsymbol{\hat{\Omega}}$, the compressed upstream magnetic field
is deemed to be wrapped in the toroidal direction perpendicular to the spin axis \citep{plot19,met19}.
This field can be considered to be globally ordered and confined within the shock plane with a preferred field orientation.
In other words, it is transverse to the local bulk velocity direction of the shocked fluid element $\boldsymbol{\hat{\beta}}$ that is identical
with the local shock normal and has a radial unit vector as $\boldsymbol{\hat{\beta}}=\boldsymbol{\hat{r}}=\boldsymbol{\hat{x}} \sin \theta \cos \phi+\boldsymbol{\hat{y}}
\sin \theta \sin \phi+\boldsymbol{\hat{z}} \cos \theta$, where $\boldsymbol{\hat{x}}$, $\boldsymbol{\hat{y}}$, $\boldsymbol{\hat{z}}$, $\theta$, and $\phi$ are mentioned below and shown in Figure \ref{fig:syn_geo}.
The field strength in the comoving frame of the shocked fluid, if the magnetization of the upstream shell $\sigma\ll 1$, is characterized by \citep{met19}
\begin{equation}
	B^{\prime}=\sqrt{64 \pi \sigma \Gamma^{2} m_{\mathrm{p}} c^{2} n_{\mathrm{ext}}}.
	\label{eq:B_p}
\end{equation}
Hereafter this section, the quantities with a prime are measured in the comoving frame of the shocked fluid.

In Figure \ref{fig:syn_geo}, two coordinate systems ($x$, $y$, $z$) and ($X$, $Y$, $Z$) are introduced with the same origin $O$ fixed in the star center.
The first one whose $z$-axis is aligned with the ejecta axis, while the second one whose $Z$-axis is aligned with the LOS ($\boldsymbol{\hat{n}}=\boldsymbol{\hat{Z}}$)
and is rotated with respect to the first one by a viewing angle of $\theta_{\rm obs}$ along the $y=Y$ axis.
For a given fluid element, its spherical coordinate forms corresponding to the systems ($x$, $y$, $z$) and ($X$, $Y$, $Z$)
are respectively ($r$, $\theta$, $\phi$) and ($r$, $\Theta$, $\Phi$),
where $r$ is the radial distance measured from the star center, $\theta$ ($\Theta$) is the polar angle (inclination angle) measured
from the ejecta axis (LOS), and $\phi$ ($\Phi$) is the azimuthal angle measured from the $x$-axis ($X$-axis). In this case, the plane of the sky is the $X$-$Y$ plane.
The ordered field direction can be parametrized such that its projection on to the $x$-$y$ plane (normal to the
ejecta axis) is $\boldsymbol{\hat{B^{\prime}}}\approx\boldsymbol{\hat{x}}\cos\phi_B+\boldsymbol{\hat{y}}\sin\phi_B$ (see the footnote 4 of Gill et al. 2020),
where $\phi_B$ is the azimuthal angle of the magnetic field that is transverse to the radial vector,
measured from the $x$-axis. When the ejecta is on-axis, i.e., $\theta_{\rm obs}=0$,
these two systems ($x$, $y$, $z$) and ($X$, $Y$, $Z$) are overlap.
This is the scenario considered in this section.

In a given fluid element, the synchrotron radiation power per unit frequency emitted by one single electron in the ordered magnetic field,
in the shocked fluid comoving frame,
is written as \citep{ryb79}
\begin{equation}
	p^{\prime}(\nu^{\prime})=\frac{\sqrt{3} e^{3} B^{\prime} \sin \vartheta_B^{\prime}}{m_{e} c^{2}} F\left(\frac{\nu^{\prime}}{\nu_{c}^{\prime}}\right),
	\label{eq:pv}
\end{equation}
where $\nu_{c}^{\prime}=\frac{3e B^{\prime} \sin \vartheta_B^{\prime} \gamma^{\prime 2}} {4 \pi m_{e} c}$ is the characteristic radiation frequency of an
electron with a thermal Lorentz factor $\gamma^{\prime}$, and $F(x)\equiv x \int_{x}^{+\infty} K_{5 / 3}(k) dk$ is the synchrotron spectrum function 
in which $K_{5 / 3}(k)$ is the modified Bessel function of the second type and $x\equiv \nu^{\prime}/\nu_c^{\prime}$ here. The pitch angle $\vartheta_B^{\prime}$ between the electron's velocity
vector and the magnetic field can be expressed by \citep[e.g.,][]{lan16}
\begin{equation}
	\sin \vartheta_B^{\prime}=\left(1-D^2\frac{\sin^2{\Theta}\cos^2{\phi_B}}{\cos^2{\Theta}+\sin^2{\Theta}\cos^2{\phi_B}}\right)^{1/2},
	\label{eq:chi_B}
\end{equation}
where $D=1/[\Gamma(1-\beta\cos\Theta)]$ is the Doppler factor and $\cos\Theta=\boldsymbol{\hat n}\cdot\boldsymbol{\hat{\beta}}$.

For a thin shell approximation to the shocked fluid, the flux density from each fluid element can be given as \citep[e.g.,][]{gran05,gill20}
\begin{equation}
	dF_{\nu}(t, \boldsymbol{\hat n})=\frac{(1+z)}{4 \pi D_{\rm L}^2} D^3 P^{\prime}(\nu^{\prime}) d\Omega,
	\label{eq:dFv}
\end{equation}
where $d\Omega=d\cos\Theta d\Phi$ is the solid angle subtended by the fluid element with respect to the central star,
$\nu^{\prime}$ relates to the observed frequency $\nu=\varepsilon/h$ with $\nu^{\prime}=(1+z)D^{-1}\nu$,
and the fluid comoving frame power of the element is
\begin{equation}
	P^{\prime}(\nu^{\prime})=\int_{\gamma_{\rm min}^{\prime}}^{\gamma_{\max}^{\prime}} N_e^{\prime}(\gamma^{\prime}) p^{\prime}(\nu^{\prime}) d \gamma,
	\label{eq:Pv}
\end{equation}
where $\gamma_{\rm min}^{\prime}$ and $\gamma_{\max}^{\prime}$ are the minimum and maximum comoving Lorentz factor of electrons, respectively.
The latter can be estimated as $\gamma_{\max}^{\prime}=(6\pi e/\sigma_{\rm T} B^{\prime})^{1/2}=10^8(B^{\prime}/1{\rm G})^{-1/2}$ by equating the electron
acceleration timescale to the synchrotron cooling timescale \citep[e.g.,][]{dai98}.
Above which the electrons in the element are assumed to be isotropic in fluid comoving frame velocity 
and to possess a hybrid thermal-nonthermal distribution in energy \citep{gia09,ress17}
\begin{equation}
	N_e^{\prime}(\gamma^{\prime})=\begin{cases}N_{e,\rm th}^{\prime}(\gamma^{\prime})=V^{\prime}K_{\rm th} \frac{\gamma^{\prime 2}}{2{\Theta_e^{\prime}}^3} e^{-\gamma^{\prime}/\Theta_e^{\prime}}, &\gamma^{\prime} \leqslant \gamma_{\rm nth}^{\prime} \\
		N_{e,\rm nth}^{\prime}(\gamma^{\prime})=V^{\prime}K_{\rm nth} {\gamma^{\prime}}^{-p_{\rm nth}}, &\gamma^{\prime}>\gamma_{\rm nth}^{\prime}, \end{cases}
	\label{eq:N_e}
\end{equation}
with $\gamma_{\rm nth}^{\prime}\simeq \frac{p_{\rm nth}-2}{p_{\rm nth}-1}\frac{m_p}{m_e} \frac{\epsilon_e \Gamma}{f_{\rm nth}}$,
where $p_{\rm nth}$ is the power-law index of nonthermal electron distribution, $\epsilon_e$ is the fraction of the shock energy goes into the electrons, 
$f_{\rm nth}$ is the fraction of electrons accelerated into the nonthermal distribution,
and while $f_{\rm th}=\left(1-f_{\rm nth}\right)$ is the fraction of electrons thermalized with temperature $\Theta_e^{\prime}\equiv kT_e^{\prime}/m_ec^2$.
The two normalization parameters can be described by 
\begin{equation}
K_{\rm th}=\frac{2 f_{\mathrm{th}} n^{\prime}}{2-e^{-y_{\rm nth}}\left(y_{\rm nth}^2+2 y_{\rm nth}+2\right)},
	\label{eq:K_th}
\end{equation}
and 
\begin{equation}
K_{\rm nth}=f_{\mathrm{nth}} n^{\prime}(p_{\rm nth}-1) {\gamma_{\rm nth}^{\prime}}^{p_{\rm nth}-1}.
	\label{eq:K_nth}
\end{equation}
Above which the particle number density $n^{\prime}$ in the fluid element behind the forward shock
can be expressed as $n^{\prime}\simeq 4\Gamma n_{\rm ext}$ \citep{blan76}, and $y_{\rm nth}\equiv \gamma_{\rm nth}^{\prime}/\Theta_e^{\prime}$ 
can be determined by solving \citep[cf. the equations (9), (17), and (18) in][]{ress17}
\begin{equation}
\frac{y_{\rm nth}^3 e^{-y_{\rm nth}}}{2-e^{-y_{\rm nth}}\left(y_{\rm nth}^2+2 y_{\rm nth}+2\right)}=\frac{f_{\rm{nth}}}{f_{\rm{th}}}(p_{\rm nth}-1).
	\label{eq:y_nth}
\end{equation}
Once $f_{\rm nth}$, $\epsilon_e$, and $\Gamma$ are given, $p_{\rm nth}$, $\gamma_{\rm nth}^{\prime}$, $y_{\rm nth}$, and $\Theta_e^{\prime}$ can then be determined.
$V^{\prime}\equiv r^2\Delta r^{\prime}=r^3/\Gamma$ is the comoving frame volume element of the shell within a unit solid angle and $\Delta r^{\prime}$ is the width of the shell.

The measured global Stokes parameters $F_{\nu}$, $Q_{\nu}$, $U_{\nu}$ are a sum over the flux $dF_{\nu}$
contributed by individual fluid elements \citep[e.g.,][]{gran03b}
\begin{equation}
	\left\{\begin{array}{l}
		U_{\nu} / F_{\nu} \\
		Q_{\nu} / F_{\nu}
	\end{array}\right\}=\left(\int dF_{\nu}\right)^{-1} \int dF_{\nu}\left\{\begin{array}{l}
		\Pi^{\prime} \sin (2 \alpha) \\
		\Pi^{\prime} \cos (2 \alpha)
	\end{array}\right\},
	\label{eq:stokes}
\end{equation}
where the PA $\alpha$, measured from the
direction of the local magnetic field, can be expressed as \citep{toma09}
\begin{equation}
	\alpha=\arctan(\tan\phi_B-\frac{\beta-\cos\Theta}{\beta\sin^2\Theta}\sin\phi_B\cos\phi_B),
	\label{eq:alpha}
\end{equation}
while $\Pi^{\prime}$ is the {\em local} degree of linear polarization from synchrotron radiation in a fluid element, given by \citep{ryb79}
\begin{equation}
	\Pi^{\prime}=\frac{\int_{\gamma_{\min}^{\prime}}^{\gamma_{\max}^{\prime}} G(x) N_e^{\prime}(\gamma^{\prime}) d \gamma^{\prime}}{\int_{\gamma_{\min}^{\prime}}^{\gamma_{\max}^{\prime}} F(x) N_e^{\prime}(\gamma^{\prime}) d \gamma^{\prime}},
\end{equation}
where $G(x)\equiv xK_{2/3}(x)$ and $F(x)$ can be referred to Equation (\ref{eq:pv}).
The integral range of the global Stokes parameters in Equation (\ref{eq:stokes}) is over the entire
ejecta surface at a fixed radius, i.e, $\Theta\in[0,\theta_{\rm ej}]$ and $\Phi$ (i.e., $\phi_B$)$\in[0,2\pi]$
for an on-axis observer \cite[e.g.,][]{gran03a,gran03b},
where $\theta_{\rm ej}$ is the half-opening angle of the ejecta.

\begin{deluxetable}{lllll}
	\label{tab:parameters}
	\tablecaption{The Best Parameter Distributions for Modeling the Spectrum of the FXRB associated with FRB 20200428D}
	\tablehead{
		\colhead{Parameters} &
		\colhead{Values}
	}
	\startdata
	\object{\bf Fixed}  &      \\
	\hline
	\object{FRB duration $\delta t$}  &  1 ms   \\
	\object{Luminosity distance $D_{\rm L}$} &  10 kpc    \\
	\object{Ejecta isotropic energy $E$} &  $10^{40}$ erg     \\
	\object{Timescale of two successive shell injection $\Delta t$}      &  $10^4$ s     \\
	\object{Shell injection rate $\dot{M}$}     &    $10^{21}~{\rm g~s^{-1}}$   \\
	\object{Shell velocity $v_{\rm w}$}     &    $0.3c$   \\
	\object{Shell magnetization $\sigma$}     &    0.01   \\
	\hline
	\object{\bf From Modeling}  &     \\
	\hline
	\object{Ratio of shock energy to electrons $\epsilon_e$}     &    $0.14^{+0.09}_{-0.08}$   \\
	\object{Non-thermal electron distribution index $p_{\rm nth}$}     &    $2.34^{+0.28}_{-0.17}$   \\
	\object{Non-thermal electron fraction $f_{\rm nth}$}     &    $0.006^{+0.003}_{-0.003}$   \\
	\object{Half-opening angle of ejecta $\theta_{\rm ej}$}     &    $0.005^{+0.001}_{-0.001}$ rad   \\
	\object{Goodness of best fitting $\chi^2/{\rm dof}$}     &  31/29    \\	
	\enddata
\end{deluxetable}

From the symmetry of $\sin(2\alpha)$, $U_{\nu}=0$. Such that the observed {\em global} degree of linear polarization from radiation of the entire ejecta is
\begin{equation}
	\begin{aligned}
		\Pi_{\nu}&=\frac{|Q_{\nu}|}{F_{\nu}}=\int D^3 \cos(2\alpha) d\Omega \int_{\gamma_{\min}^{\prime}}^{\gamma_{\max}^{\prime}} B^{\prime} \sin\vartheta_B^{\prime} G(x) N^{\prime}_e(\gamma^{\prime}) d\gamma^{\prime} \\
		&\times\left[\int D^3 d\Omega \int_{\gamma_{\min}^{\prime}}^{\gamma_{\max}^{\prime}} B^{\prime} \sin\vartheta_B^{\prime} F(x) N^{\prime}_e(\gamma^{\prime}) d\gamma^{\prime}\right]^{-1},
		\label{eq:Pi_nu}
	\end{aligned}
\end{equation}
and its corresponding PA is
\begin{equation}
	\chi_{\rm p}=\frac{1}{2}\arctan{\left(\frac{U_{\nu}}{Q_{\nu}}\right)}=0
	\label{eq:chi_p}
\end{equation}

\begin{figure}[ht]
	\includegraphics[width=\columnwidth]{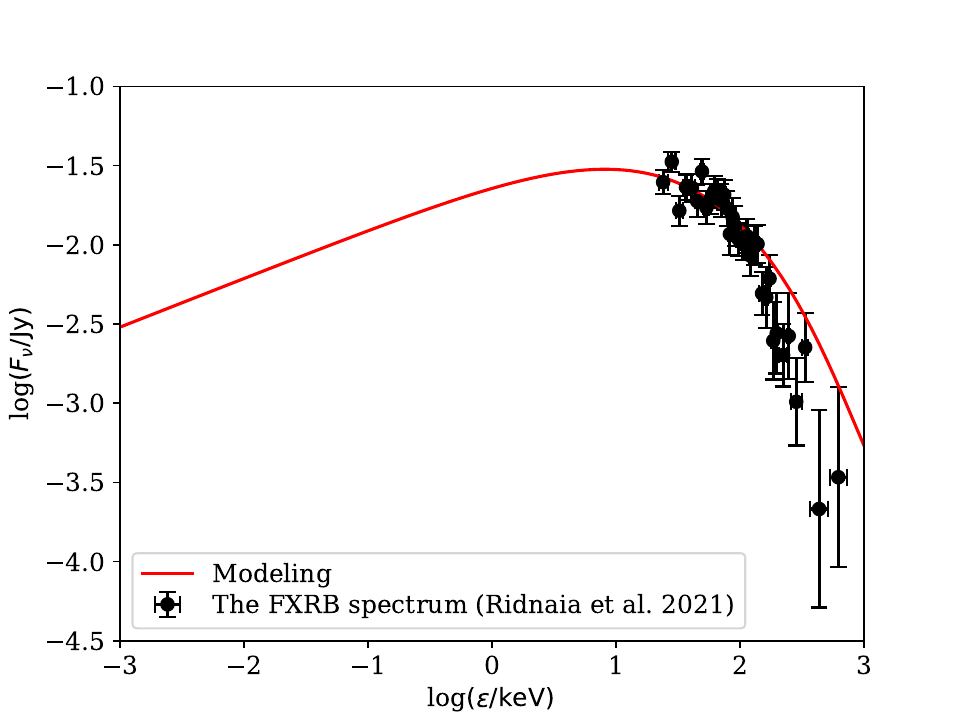}
	\includegraphics[width=\columnwidth]{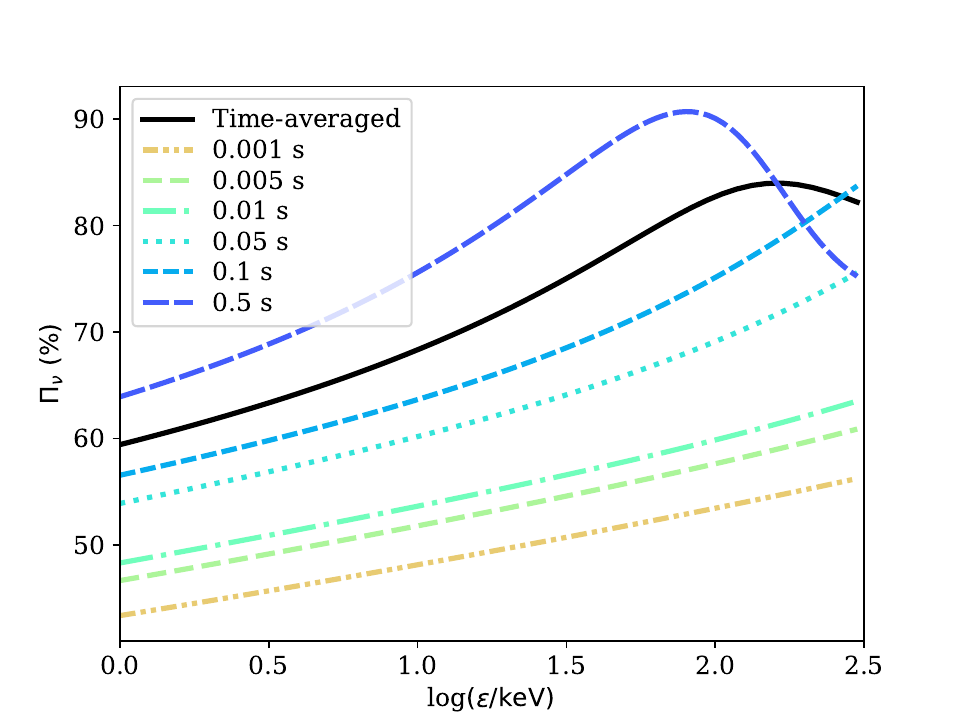}
	\includegraphics[width=\columnwidth]{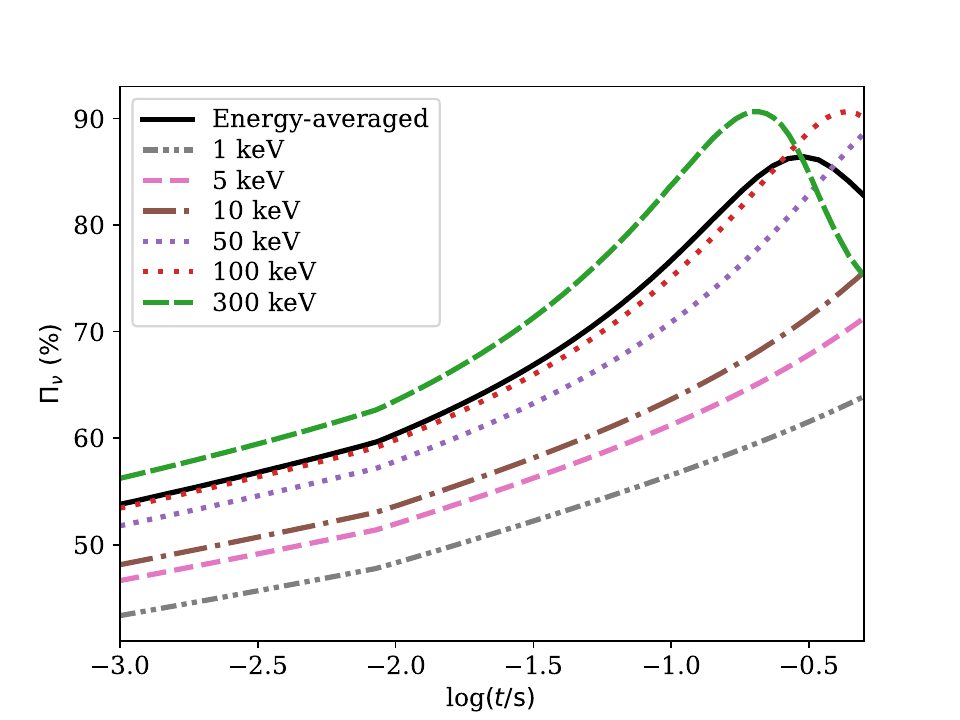}
	\caption{{\it Top}: The best spectral modeling for the FXRB associated with FRB 20200428D.
		{\it Middle}: Time-sliced and averaged PDs in the energy range of $1-300$ keV.
		{\it Bottom}: Energy-sliced and averaged PDs in the time range of $10^{-3}-0.5$ s since $t_0-0.220$ s in which $t_0=$ 14:34:24.447 UT is the trigger time \citep{rid21}.}
	\label{fig:spec-pd}
\end{figure}

\subsection{Results}
\label{subsec:results3}
We combine the dynamics of relativistic shock in Section \ref{subsec:dynamics} and the spectrum calculation from Equation (\ref{eq:stokes}) 
to model the FXRB associated with FRB 20200428D.
The adopted values of the model parameters include those from the observations for FRB 20200428D 
such as its duration $\delta t\sim1$ ms \citep{chime20b} and distance $D_{\rm L}\sim10~{\rm kpc}$ \citep{zhong20}, 
and those describe the properties of the relativistic ejecta and shell that fulfill the FRB creation such as the injection timescale $\Delta t\sim10^4$ s, injection rate $\dot{M}\sim10^{21}~{\rm g~s^{-1}}$, velocity $v_{\rm w}\sim0.3c$, and magnetization $\sigma\sim0.01$ of the shell, as well as the ejecta isotropic energy $E\sim E_{\rm FXRB}\sim10^{40}$ erg \citep{wu20}. 
Moreover, we take four parameters $\epsilon_e$, $p_{\rm nth}$, $f_{\rm nth}$, and $\theta_{\rm ej}$ as free. By the MCMC algorithm, we obtain the best-fit values for parameters $\epsilon_e=0.14^{+0.09}_{-0.08}$, $p_{\rm nth}=2.34^{+0.28}_{-0.17}$, $f_{\rm nth}=0.006^{+0.003}_{-0.003}$, and $\theta_{\rm ej}=0.005^{+0.001}_{-0.001}$ rad, listed in Table \ref{tab:parameters}. The best-fit result for the FXRB spectrum is plotted in the top panel of Figure \ref{fig:spec-pd}. 
From the values of $\theta_{\rm ej}$ and $f_{\rm nth}$, one can see that the ejecta has a very narrow half-opening angle and the FXRB spectrum is thermal-dominated, i.e., a relativistic Maxwellian distribution (see the first line of Equation (\ref{eq:N_e})).

By these parameter values, one can numerically calculate the time-sliced and averaged PDs in the energy range $1-300$ keV as well as the energy-sliced and averaged PDs in the FXRB time range $10^{-3}-0.5$ s, see the middle and bottom panels of Figure \ref{fig:spec-pd}. 
From the results, one can find that the PDs are relatively high, falling into the range of $40\%-90\%$, even they are time-dependent and energy-dependent.
Moreover, the PDs are larger in high energy band than in low energy band on average, 
which are exactly opposite from the PD results within the model of the emission of a trapped fireball modified by RCS.
In addition, the PAs are flat during the FXRB duration and are independent of energy as noticed previously.

\section{Summary and Discussion}
\label{sec:summary}
We have summarized three commonly-mentioned models for FXRB origin: the emission of a trapped fireball modified by RCS, the polar trapped-expanding fireball, and the synchrotron radiation in a far-away relativistic forward shock, then discussed whether they can address the issue over the discrepancy between the FXRB and OXRBs in SGR 1935.
In order to identify the realistic origin of an FXRB, we have presented an investigation for possible X-ray polarization characteristics of the FXRB associated with an FRB like FRB 20200428D within these three models and expected future X-ray polarization observations from telescopes, though current and forthcoming X-ray polarimeters such as IXPE \citep{wei22}, eXTP \citep{san19}, and POLAR-2 \citep{pro23} are very far from allowing one to measure polarization variations in both energy and time for the events discussed here.
The main polarization results are obtained as follows:
\begin{itemize}
	\item If the FXRB is produced by the emission of a trapped fireball modified by RCS, its PDs usually vary with energy, rotation phase, and $\chi$ as well as $\varsigma$ that the spin axis makes with the LOS as well as magnetic axis. Moreover, the PDs are averagely smaller in high energy band (e.g., 50-300 keV) than in low energy band (e.g., 1-50 keV) for a fixed $\chi$ as well as $\varsigma$. While for the PAs, they are usually dependent on 
	rotation phase and $\chi$ as well as $\varsigma$, but independent of energy. 
	\item If the FXRB is created by a polar trapped-expanding fireball, its PDs are usually nearly invariant with rotation phase and $\chi$ as well as $\varsigma$, but vary with energy. In contrast, its PAs usually vary with rotation phase and $\chi$ as well as $\varsigma$, but not energy. 
	\item If the FXRB is generated by the synchrotron radiation in a far-away relativistic shock, its PDs vary with dynamical time and energy, but its PAs only have a constant value and do not vary with either dynamical time or energy. Moreover, the PDs are larger in high energy band than in low energy band on average, 
	which are exactly opposite from the PD results within the model of the emission of a trapped fireball modified by RCS.
\end{itemize}

The differences of polarization (both PD and PA) variations with phase/time and energy among these three models can be used to diagnose the origin of an FXRB. 
For instance, looking at PA variations with phase/time and energy,
if an FXRB is generated by the synchrotron radiation in a relativistic shock far outside magnetosphere, it only has a constant PA value with time and energy, 
which is analogous to flat PA curves in some FRBs such as FRBs 20121102A \citep{gaj18,mich18}, 20180916B \citep{chime19,cha20,nim21,pas21,sand22}, 
20190711A \citep{day20,kumar21}, 20190303A as well as 20190417A \citep{feng22}, and 20190604A \citep{fon20}. 
Otherwise, 
if the FXRB is produced by the emission from a trapped fireball modified by RCS or a polar trapped-expanding fireball in magnetosphere, 
in addition to possibly having a constant PA value with phase, 
it is more likely to have a variable PA during its short duration, somewhat like a few FRBs with varying PA swings in phase, 
e.g., FRBs 20180301A \citep{luo20} and 20201124A \citep{xu22,kumar22}. 
Furthermore, if the FXRB has some cuts with rotation phase observed by future X-ray polarimeters, 
it is more likely created by a polar trapped-expanding fireball than the emission from a trapped fireball modified by RCS.

The extraordinary features of the FXRB associated with FRB 20200428D prefer a polar locale interpretation for both the FRB and its FXRB \citep{younes20b,younes21,zhu23}. This interpretation seems to imply periodic bursts for a repeating FRB source theoretically. The bursts in most of repeating sources, inversely, appear in random phases observationally. This dichotomy may shed light on an outburst-driven evolving complex magnetic field topology \citep{younes20b} such as multipolar field in an active young magnetar, 
where bursts and their FXRBs are both produced at magnetic poles.

\acknowledgments
We are grateful to the referee for helpful comments and suggestions.
S.Q.Z. is very grateful to Can-Min Deng for his encouragement to this work.
This work is supported by the starting Foundation of Guangxi University of Science and Technology (grant No. 24Z17).
L.L. is supported by the National Natural Science Foundation of China (grant No. 12303050).
Z.G.D. is supported by the National SKA Program of China (grant No. 2020SKA0120302) 
and National Natural Science Foundation of China (grant No. 12393812).

\appendix
\section{Parameters in a Trapped-Expanding Fireball}
The values/expressions of $\delta^{\prime}$, $\zeta^{\prime}$, and $\tau_{ \pm 0}$ are \citep{wada23}
\begin{equation}
	\delta^{\prime}=\begin{cases} -17/4, & {\rm RD/O{\text -}mode/lL}  \\
		                                                    -11/4, & {\rm RD/O{\text -}mode/hL}  \\
		                                                     -5/4, &  {\rm RD/E{\text -}mode/lL}, 
	                                                     \end{cases}
	\label{eq:delta_p}
\end{equation}

\begin{equation}
	\zeta^{\prime}=\begin{cases} -1/4, & {\rm RD/O{\text -}mode/lL}  \\
		5/4, & {\rm RD/O{\text -}mode/hL}  \\
		11/4, &  {\rm RD/E{\text -}mode/lL}, 
	\end{cases}
	\label{eq:zeta_p}
\end{equation}
and
\begin{equation}
	\tau_{ \pm 0}=\begin{cases} n_{\pm}(T_0,B_0)\sigma_{\rm T}r_0, & {\rm RD/O{\text -}mode/lL}  \\
		n_{\pm}(T_0)\sigma_{\rm T}r_0, & {\rm RD/O{\text -}mode/hL}  \\
		(4\pi^2T_0^2B_0^{-2}/5) n_{\pm}(T_0,B_0)\sigma_{\rm T}r_0, &  {\rm RD/E{\text -}mode/lL}, 
	\end{cases}
	\label{eq:tau_pm0}
\end{equation}


\end{document}